\documentclass[aps,twocolumn,superscriptaddress]{revtex4-2}

\usepackage{graphicx}
\usepackage{amssymb,amsmath}
\usepackage{braket}
\usepackage{placeins}
\usepackage{natbib}
\usepackage{xr}
\usepackage{xcolor}
\usepackage{soul}
\usepackage[colorlinks=true, citecolor={blue!80!black}, urlcolor={blue!80!black}, linkcolor = {blue!80!black}]{hyperref}

\makeatletter
\newcommand*{\addFileDependency}[1]{
    \typeout{(#1)}
    \@addtofilelist{#1}
    \IfFileExists{#1}{}{\typeout{No file #1.}}
}
\makeatother

\newcommand*{\myexternaldocument}[1]{%
    \externaldocument{#1}%
    \addFileDependency{#1.aux}%
}

\myexternaldocument{supprefs}

\begin{document}

\title{A Room-Temperature Solid-State Maser Amplifier}

\author{Tom Day}
\affiliation{School of Electrical Engineering and Telecommunications, UNSW Sydney, Sydney, NSW 2052, Australia}
\author{Maya Isarov}
\affiliation{School of Electrical Engineering and Telecommunications, UNSW Sydney, Sydney, NSW 2052, Australia}
\author{William J. Pappas}
\affiliation{School of Physics, UNSW Sydney, Sydney, NSW 2052, Australia}
\author{Brett C. Johnson}
\affiliation{School of Science, RMIT University, Melbourne, Victoria 3001, Australia}
\author{Hiroshi Abe}
\affiliation{National Institutes for Quantum Science and Technology (QST), 1233 Watanuki, Takasaki, Gunma 370-1292, Japan}
\author{Takeshi Ohshima}
\affiliation{National Institutes for Quantum Science and Technology (QST), 1233 Watanuki, Takasaki, Gunma 370-1292, Japan}
\affiliation{Department of Materials Science, Tohoku University, Aoba, Sendai, Miyagi 980-8579, Japan}
\author{Dane R. McCamey}
\affiliation{School of Physics, UNSW Sydney, Sydney, NSW 2052, Australia}
\author{Arne Laucht}
\affiliation{School of Electrical Engineering and Telecommunications, UNSW Sydney, Sydney, NSW 2052, Australia}
\author{Jarryd J. Pla}
\email[]{jarryd@unsw.edu.au}
\affiliation{School of Electrical Engineering and Telecommunications, UNSW Sydney, Sydney, NSW 2052, Australia}
\date{\today}

\begin{abstract}
Masers once represented the state-of-the-art in low noise microwave amplification technology, but eventually became obsolete due to their need for cryogenic cooling. Masers based on solid-state spin systems perform most effectively as amplifiers, since they provide a large density of spins and can therefore operate at relatively high powers. Whilst solid-state masers oscillators have been demonstrated at room temperature, continuous-wave amplification in these systems has only ever been realized at cryogenic temperatures. Here we report on a continuous-wave solid-state maser amplifier operating at room temperature. We achieve this feat using a practical setup that includes an ensemble of nitrogen-vacancy center spins in a diamond crystal, a strong permanent magnet and simple laser diode. We describe important amplifier characteristics including gain, bandwidth, compression power and noise temperature and discuss the prospects of realizing a room-temperature near-quantum-noise-limited amplifier with this system. Finally, we show that in a different mode of operation the spins can be used to reduce the microwave noise in an external circuit to cryogenic levels, all without the requirement for physical cooling.
\end{abstract}

\maketitle

\section{Introduction}\label{sec:intro}
The detection of weak microwave signals is a challenge that lies at the heart of many modern technologies, including deep-space satellite communication systems \cite{macgregor2008low}, radio telescopes \cite{wilson2009tools}, radar and electron spin resonance (ESR) spectrometers \cite{bienfait2016reaching}. The Voyager 1 space probe transmits microwave signals at a power of approximately 20~W, reducing to a mere $10^{-18}$~W of power by the time the transmissions reach Earth \cite{yuen2013deep}. In order to detect such faint signals, microwave receivers must add as little noise as possible, taking advantage of ultra-low noise amplifiers to first boost the signals before measurement. 

Maser amplifiers are devices that exploit stimulated emission in an inverted ensemble of microwave-frequency emitters -- often in the form of paramagnetic centers such as spins -- to achieve low-noise amplification of microwave signals. The noise temperature of a maser amplifier $T_{\rm m}$, which quantifies the amount of noise added to a signal before it is amplified, can theoretically reach the quantum limit of $T_{\rm m}\approx\hbar\omega_{\rm s}/k_{\rm B}$, where $\omega_s$ is the spin transition frequency and $\hbar$ and $k_{\rm B}$ are the reduced Planck and Boltzmann constants, respectively. For a maser operating in the microwave X-band ($\omega_{\rm s}/2\pi\approx 10$~GHz), this temperature can be as low as $T_{\rm m} = 0.48$~K.

Maser amplifiers based on solid-state spin systems such as ruby \cite{makhov1958maser} once represented the gold-standard in low-noise microwave amplification technology. The requirement for solid-state maser amplifiers to be cooled to cryogenic temperatures (typically $\lesssim 4.2$~K) \cite{siegman1964microwave} saw these systems eventually replaced by modern transistor-based amplifiers \cite{joshin1983noise}. However, pioneering experiments using organic \cite{oxborrow2012room} or diamond-based \cite{breeze2018continuous} spin systems have demonstrated that solid-state masers can operate at room temperature, generating a resurgence in their interest. Due to the tantalizing prospect of achieving quantum-limited noise performance, there has been a renewed effort \cite{fischer2018highly,sherman2021performance,sherman2022diamond,wang2023tailoring,gottscholl2023room} to develop room-temperature solid-state maser amplifiers. 

Maser amplification of short ($\sim 30~\mu$s) pulsed microwave signals was recently observed at room temperature using a spin system in an organic host \cite{wang2023tailoring}, whilst a continuously operating diamond maser amplifier was realized at liquid nitrogen temperatures ($\sim 77$~K) \cite{sherman2022diamond}. However, the demonstration of a widely usable continuous-wave maser amplifier operating under ambient conditions remains a highly desirable objective.

\begin{figure*}[ht!]
\includegraphics[width=2\columnwidth]{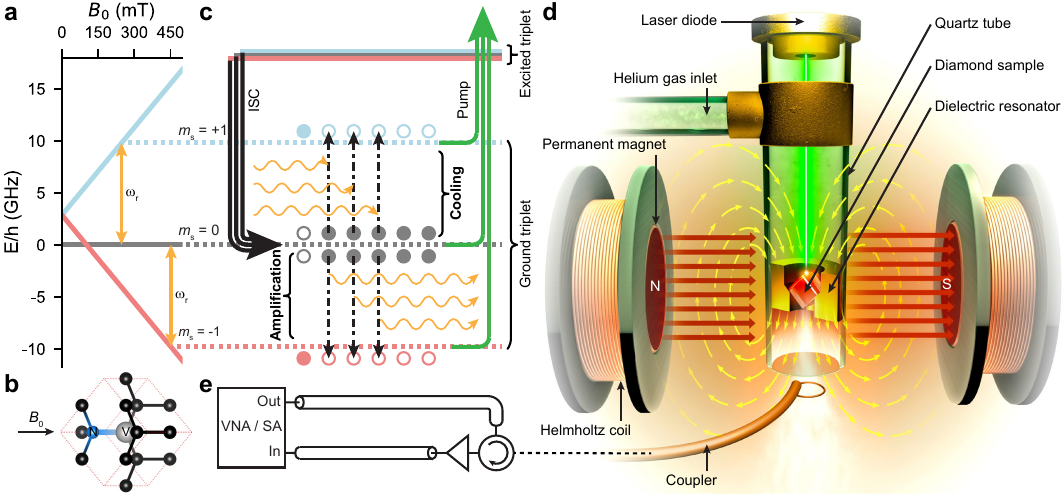}
\caption{\textbf{Configuration of the NV diamond room-temperature maser amplifier/cooler.} \textbf{a}, Energy level structure of the NV$^-$ triplet ground state, with the $B_{\rm 0}$ magnetic field applied parallel to the NV$^-$ axis. \textbf{b}, Crystal structure of a single NV$^-$ center in diamond created using the VESTA software package \cite{momma2011vesta}, showing the NV$^-$ axis aligned with the applied magnetic field $B_{\rm 0}$. \textbf{c}, Diagram of the absorptive cooling transition (upper) and emissive masing transition (lower). Pumping with green light causes the spins to enter an excited state, where they preferentially relax to the $m_{\rm s}=0$ spin state via an intersystem crossing (ISC). \textbf{d}, An artist's impression of the maser amplifier/cooler setup. The quartz tube with dielectric resonator and diamond assembly are inserted inside a high conductivity copper cavity (not shown) with a fundamental resonance frequency of $\sim 11$~GHz, to suppress any radiation loss from the dielectric resonator. \textbf{e}, Schematic of the microwave setup used to probe the system in reflection with a vector network analyzer (VNA) or measure noise using a spectrum analyzer (SA). The loop coupler is connected to a low-loss circulator, which is attached to the output of the VNA at one port and via a low-noise microwave amplifier to the input of the VNA or SA at the other port.}
\label{fig:fig1}
\end{figure*}

In this work we describe a continuous-wave maser amplifier based on a solid-state system that operates at room temperature. We use an ensemble of nitrogen-vacancy (NV) center spins in a bulk diamond crystal as the gain medium and couple this to a high quality factor microwave dielectric resonator. We achieve gains as high as 30~dB and a gain-bandwidth-product of up to 4.5~MHz. Our analysis shows that the intrinsic noise temperature of the spin gain medium is close to the quantum limit, whilst the dominant source of noise comes from resonator losses. We further demonstrate that the same device can be used to remove microwave noise in a circuit \cite{wu2021bench,ng2021quasi,zhang2022microwave,fahey2023steady,blank2023anti} by acting like a matched load with an effective microwave temperature of 66~K, without the need for any cryogenic cooling. We develop a modern quantum optics theoretical model \cite{gardiner1985input} to describe the maser amplifier and microwave cooler system and obtain quantitative agreement with our measurement results. Finally, we discuss relatively straight-forward improvements that can be made to our system to push its noise performance towards the quantum limit. These results demonstrate that NV spin ensembles in diamond coupled to microwave resonators constitute a system with exceptional promise for performing ultra-low noise detection of microwave signals at room temperature.

\section{Results}
\subsection{Maser amplifier design}\label{sec:design}
Solid-state masers have been realized at room temperature using several different physical spin systems, including organic crystals of pentacene doped in para-terphenyl (Pc:PTP) \cite{oxborrow2012room} or 6,13-diazapentacene-doped para-terphenyl (DAP:PTP) \cite{ng2023move} and diamonds containing negatively charged nitrogen vacancy (NV$^-$) centers ~\cite{jin2015proposal,breeze2018continuous,zollitsch2023maser}. In each case the spin system was coupled to a high quality factor microwave dielectric resonator and population inversion of the spins was achieved via optical pumping using moderate \cite{breeze2018continuous} or high power \cite{oxborrow2012room} lasers. Diamond is an ideal material for hosting the gain medium due to its high thermal conductivity (2,200 Wm$^{-1}$K$^{-1}$ at 294~K) \cite{graebner1995thermal}, which allows any heat generated by the optical pumping to be rapidly dissipated in its surroundings.

In this work we utilize a high concentration NV$^-$ diamond bulk crystal as the gain medium. The energy level structure of the NV$^-$ triplet ground state is shown in Fig.~\ref{fig:fig1}a. When a static magnetic field $B_{\rm 0}$ is applied parallel to one of the four equivalent NV $\langle 111\rangle$ directions (see Fig.~\ref{fig:fig1}b), the $m_{\rm s} = \pm 1$ states of the NV$^-$ spins Zeeman split with a gradient of $\pm \gamma_{\rm e}$ (with $\gamma_{\rm e}/2\pi = 28$~GHz/T), offset by a zero field splitting of $D/2\pi = 2.87$~GHz \cite{doherty2012theory,barry2020sensitivity}. The NV$^-$ spins can be readily polarized into the ground triplet $m_{\rm s} = 0$ state by optical pumping using green ($\sim$ 520 nm) light. At a sufficiently large $B_{\rm 0}$, the $m_{\rm s} = -1$ spin state energy falls below that of the the $m_{\rm s} = 0$ state and the optical pumping results in a population inversion \cite{breeze2018continuous} (see Fig.~\ref{fig:fig1}c, lower), fulfilling a key condition for implementing a maser amplifier. For smaller $B_{\rm 0}$ fields where the $m_{\rm s} = 0$ state is lowest in energy (see Fig.~\ref{fig:fig1}c, upper), the NV$^-$ spins absorb microwaves and can thus be used to reduce the noise present in a circuit \cite{wu2021bench,ng2021quasi,zhang2022microwave,fahey2023steady,blank2023anti}.

Figure~\ref{fig:fig1}d depicts the setup utilized in our work. The diamond sample employed is a rectangular prism of dimensions 1.8~mm $\times$ 1.9~mm $\times$ 2.0~mm and is grown by chemical vapor deposition (CVD) with $\{100\}$ faces. It has a nominal concentration of $\sim 3$~ppm NV$^-$, as specified by the manufacturer (see Methods). The diamond prism is positioned inside a cylindrical dielectric resonator that is custom-made from a high-permittivity ($\epsilon_{\rm r}\approx30$) ceramic material, which exhibits a fundamental microwave mode in the microwave X-band at $\omega_{\rm r}/2\pi=9.8$~GHz (see Supplementary Information, Section~\ref{sup:subsec:DR} for details). The resonator contains a sloped inner lip to support the diamond and ensure that the $B_{\rm 0}$ field can be aligned along one of the $\langle 111\rangle$ directions of the crystal. The high permittivity of the ceramic dielectric material reduces the resonator volume for a given frequency and thus enhances the interaction strength between the mode and the spins, which is particularly important when the diamond sample is smaller than the volume of the mode's magnetic field (see Supplementary Information, Section~\ref{sup:subsec:DR}). An adjustable inductive loop (see Methods) is used to couple the resonator to external circuitry (see Fig.~\ref{fig:fig1}e), with an external coupling quality factor ($Q_{\rm e}$) that can be made smaller or larger than the resonator's internal quality factor ($Q_{\rm i}$), which has a nominal value of $Q_{\rm i} \approx 11,000$.

To access the maser amplification regime using the NV system, a $B_{\rm 0}$ field of approximately 450~mT must be applied to bring the $m_{\rm s} = 0$ to $m_{\rm s} = -1$ spin transition frequency to $\omega_{\rm r}/2\pi=9.8$~GHz. The generation of such a strong magnetic field is typically achieved using large and high power consumption electromagnets \cite{breeze2018continuous,zollitsch2023maser}, which present challenges for the ultimate goal of creating a low size, weight and power amplifier. In this work, the magnetic field is provided by a permanent magnet, with a small electromagnetic Helmholtz sweep coil to allow fine tuning of the field (see Methods). The green pump laser is supplied by a simple 1.5~W optical output power laser diode, which is less expensive, has a relatively low power consumption and a small form factor compared to other pump lasers. 

Our ceramic resonator has a low thermal conductivity (2.56 Wm$^{-1}$K$^{-1}$ at 294~K), which prevents adequate heat dissipation and causes both the diamond and resonator temperature to rise during optical pumping. Since the spin relaxation time $T_1$ (and therefore our ability to optically pump the NV$^-$ spins) depends strongly on temperature \cite{zollitsch2023maser}, maintaining the diamond sample at room temperature is vital. Further, increasing the resonator temperature leads to higher levels of thermal noise, which is also to be avoided. To better thermalize the diamond and resonator assembly, we flow room-temperature helium gas -- which has a thermal conductivity of 0.15 Wm$^{-1}$K$^{-1}$ at 294~K, approximately a factor $6\times$ greater than air -- continuously over the sample (see Fig.~\ref{fig:fig1}a). This maintains both the resonator and diamond close to room temperature, even when operating the laser diode at full power, as verified by probing the resonator noise and performing spin $T_1$ measurements (see Supplementary Information, Sections~\ref{sup:subsec:He} and \ref{sup:subsec:T1T2}). With helium gas applied, the spin $T_1$ time for our sample is $T_1 = 6.7(1)$~ms (see Supplementary Information, Section~\ref{sup:subsec:T1T2}).

\subsection{Input-output model} 
The theory describing gain and noise in maser and other types of negative resistance amplifiers was established over six decades ago \cite{siegman1964microwave}. These theories use a semi-classical wave-approach analysis based on simple thermodynamic arguments. Here we develop a quantum optics theoretical model for our system which takes into account microscopic details of the spin gain medium and relates important quantities like gain and noise to modern circuit quantum electrodynamics parameters \cite{blais2021circuit}. In doing so, we are able to describe the full frequency dependence of the reflection and noise spectra when operating the system in either the maser amplification or microwave cooling regimes. In Sections~\ref{subsec:gain} and \ref{subsec:noise} we show that the model reproduces key predictions of the semi-classical theory. Our model is based on the input-output formalism of Gardiner and Collett \cite{gardiner1985input} and builds on previous input-output descriptions of resonators coupled with inverted \cite{julsgaard2012dynamical} or non-inverted \cite{kurucz2011spectroscopic,diniz2011strongly,fahey2023steady, bienfait2017magnetic} spin ensembles. A full derivation of the model is presented in the Supplementary Information (Section~\ref{sup:sec:Theory}), below we summarize the key points and predictions.

Without damping, the system can be described by the following Hamiltonian:
\begin{align}\label{eqn:H}
    H &= \hbar \omega_{\rm r} a^\dagger a + \frac{\hbar}{2} \sum_{j=1}^N \omega_j \sigma_{\rm z}^{(j)} + \hbar \sum_{j=1}^{N} \left( g_j^* \sigma_{\rm +}^{(j)} a + g_j \sigma_{\rm -}^{(j)} a^\dagger \right),
\end{align}
where $a$ ($a^\dagger$) is the annihilation (creation) operator for the resonator mode, $\sigma_{\rm z}^{(j)}$, $\sigma_{\rm +}^{(j)}$ and $\sigma_{\rm -}^{(j)}$ are, respectively, the Pauli Z, raising and lowering operators describing the $j$-th spin in the ensemble (out of $N$ total spins), which has a resonance frequency of $\omega_j$ and is coupled with a strength $g_j$ to the resonator. The distribution of spin frequencies $\omega_j$ is free to take any form here, with Lorentzian and Gaussian profiles being common.

\begin{figure}[t!]
\includegraphics[width=\columnwidth]{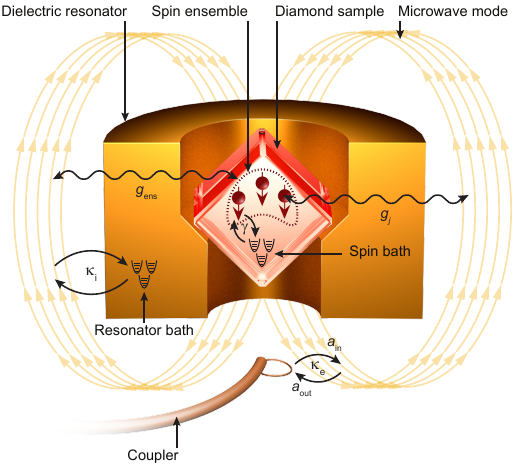}
\caption{\textbf{Input-output system interactions.} Diagram showing the dielectric resonator and diamond assembly together with the loop coupler, which carries input and output fields $a_{\rm in}$ and $a_{\rm out}$, respectively. The input/output field operators are coupled to the resonator at a rate $\kappa_{\rm e}$, whilst the resonator field is coupled to a bath (representing internal channels of loss) at a rate $\kappa_{\rm i}$. The diamond sample contains an ensemble of spins, which are individually coupled to the resonator field with a strength $g_j$ and coupled as an ensemble to the resonator at an enhanced rate $g_{\rm ens}$. Each spin couples to an independent bath at a rate $\gamma$.}
\label{fig:fig2}
\end{figure}

Damping is introduced by coupling the resonator mode $a$ to (i) bath modes in the external cable, which carries input $a_{\rm in}$ and output $a_{\rm out}$ fields that couple to the resonator field at a rate $\kappa_{\rm e} = \omega_{\rm r}/Q_{\rm e}$, and (ii) bath modes representing any internal sources of resonator loss, which are coupled to the resonator field at a rate $\kappa_{\rm i} = \omega_{\rm r}/Q_{\rm i}$. Each spin is assumed to couple to an independent bath at a rate $\gamma$, which is at the effective spin temperature $T_{\rm s}$. We note that $T_{\rm s}$ is negative for an inverted spin ensemble and positive for a non-inverted ensemble (see Supplementary Information, Section~\ref{sup:subsec:baths}), relevant for the maser amplification and microwave cooling regimes, respectively. Figure~\ref{fig:fig2} presents a summary of the various elements and interactions present in the system, as well as the sources of damping. 

When probing the system with a microwave tone $\langle a_{\rm in}\rangle$ and measuring the reflected signal $\langle a_{\rm out}\rangle$ via the loop coupler (see Figs.~\ref{fig:fig1}d,e), the model predicts a reflection coefficient:
\begin{equation} \label{eqn:r}
    r^\pm = \frac{\langle a_{\rm out}\rangle}{\langle a_{\rm in}\rangle}=\frac{i\kappa_{\rm e}}{\omega - \omega_{\rm r} + i\bar{\kappa} \pm K\left(\omega\right)} -1,
\end{equation}
where $\bar{\kappa}=(\kappa_{\rm e} + \kappa_{\rm i})/2$, $K\left(\omega\right)$ is a function that contains information about the spin resonance frequency distribution and their coupling to the resonator \cite{kurucz2011spectroscopic,grezes2016towards}, and the $\pm$ in the denominator indicates the solution for an inverted ($+$) or non-inverted ($-$) ensemble. For a spin ensemble with a Lorentzian distribution of resonance frequencies $\omega_j$, one can show \cite{kurucz2011spectroscopic,diniz2011strongly}:
\begin{equation}
    K(\omega) = \frac{g_{\rm ens}^2}{(\omega-\omega_{\rm s})+i(\Gamma+\gamma)/2},\label{eqn:kw}
\end{equation}
where $g_{\rm ens} = (\sum_j{|g_j|^2})^{1/2}$ is the collective coupling strength of the spin ensemble to the resonator, $\omega_{\rm s}$ is the average spin resonance frequency and $\Gamma$ is the characteristic width of the spin frequency distribution. In the Supplementary Information (Section~\ref{sup:sec:kc}) we also present an analytical expression of $K\left(\omega\right)$ for a Gaussian spin distribution, however, $K(\omega)$ can be numerically evaluated for any profile.

We are also able to predict the output noise spectrum of the system:
\begin{align}\label{eqn:nout}
    n_{\rm out}^\pm = R_{\rm in}^\pm\left(n_{\rm in} + \frac{1}{2}\right) +
    R_{\rm s}^\pm\left(n_{\rm s} + \frac{1}{2}\right) +
    R_{\rm i}^\pm\left(n_{\rm i} + \frac{1}{2}\right),
\end{align}
with,
\begin{equation}
\begin{aligned}\label{eqn:R}
    R_{\rm in}^\pm &= |r^\pm|^2,\\
    R_{\rm s}^\pm &= \frac{\kappa_{\rm e}}{\left|\omega - \omega_{\rm r} + i\bar{\kappa} \pm K\left(\omega\right)\right|^2}C(\omega),\\
    R_{\rm i}^\pm &= \frac{\kappa_{\rm e}\kappa_{\rm i}}{\left|\omega - \omega_{\rm r} + i\bar{\kappa} \pm K\left(\omega\right)\right|^2},
\end{aligned}
\end{equation}
where $n_{\rm out}^\pm$ is the frequency-dependent number of noise photons in the output field $a_{\rm out}$ (i.e., traveling away from the resonator via the coupler), $n_{\rm in}$ is the number of noise photons in the input field $a_{\rm in}$ (traveling to the resonator via the coupler), $n_{\rm s}$ is the number of noise photons in the spin bath, $n_{\rm i}$ is the number of noise photons in the resonator loss bath and $C(\omega)$ is a convolution function that depends on the details of the spin distribution (see Supplementary Information, Section~\ref{sup:sec:kc}). We assume that the various baths are all in thermal states, with thermal photon populations given by the Bose-Einstein distribution $n = 1/(\exp[\hbar\omega/(k_BT)]-1)$, where $T$ is the effective temperature of the bath (see Supplementary Information, Section~\ref{sup:subsec:baths} for details). Finally, we note that the $1/2$ terms in Eq.~\ref{eqn:nout} correspond to the quantum limit of noise (vacuum fluctuations) introduced by each bath.

\subsection{Maser amplifier performance}\label{sec:amplification}
\begin{figure*}[ht!]
\includegraphics[width=2\columnwidth]{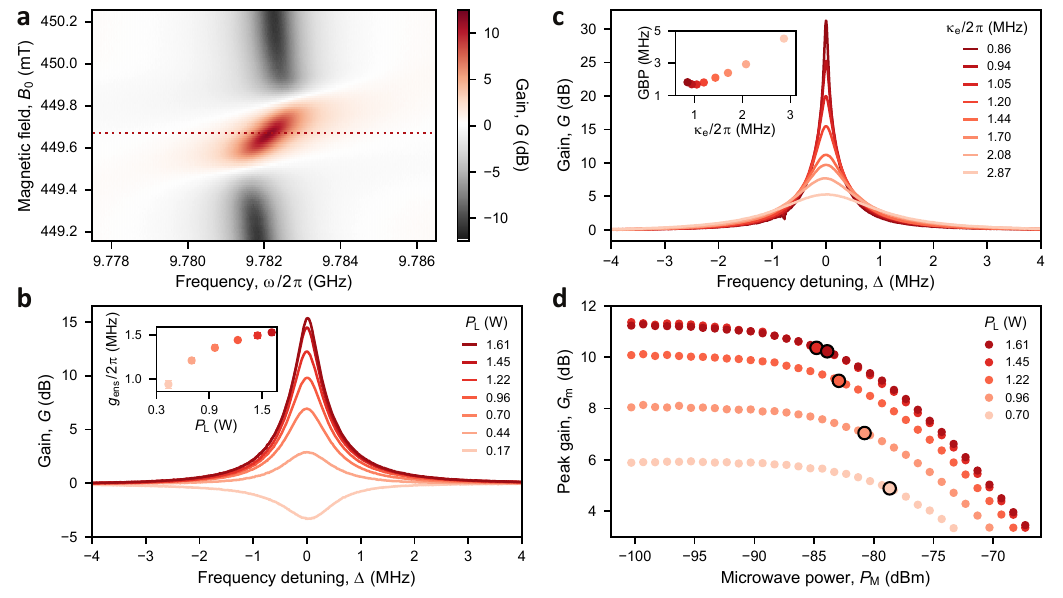}
\caption{\textbf{Maser amplifier gain measurements.} \textbf{a}, Power gain $G$ measured as $B_{\rm 0}$ is swept across the NV$^-$ spin amplification transition. The dotted line represents a line-cut at which the measurements in panels b and c are taken. \textbf{b}, Amplifier gain (at the line-cut shown in panel a) as a function of laser power $P_{\rm L}$, plot against the detuning $\Delta$ from the center frequency $\omega/2\pi = 9.782$~GHz. Inset shows the ensemble coupling strength $g_{\rm ens}$ against $P_{\rm L}$, extracted from fits of the gain curves to our theory. \textbf{c}, Amplifier gain (at the line-cut shown in panel a) as a function of the external coupling $\kappa_{\rm e}$. Inset presents the GBP extracted from the gain curves as a function of $\kappa_{\rm e}$. \textbf{d}, Peak amplifier gain $G_{\rm m}$ as a function of the applied input microwave power $P_{\rm M}$ taken at several different laser powers. The larger dots indicate the 1~dB compression points $P_{\rm 1dB}$ at each $P_{\rm L}$. }
\label{fig:fig3}
\end{figure*}

\subsubsection{Gain}\label{subsec:gain}
We proceed to operate the system as a maser amplifier by pumping the NV$^-$ spins with the laser diode at a power of $P_{\rm L} = 1.61$~W, whilst simultaneously probing the resonator using a vector network analyzer (VNA), allowing us to measure the reflection coefficient $r^+$ (Eq.~\ref{eqn:r}). We adjust the position of the permanent magnet poles to provide a field close to $B_{\rm 0} \approx 450$~mT, which brings the $m_{\rm s} = -1$ to $m_s = 0$ spin transition near resonance with the dielectric resonator. We perform a fine sweep of the field using the Helmholtz coil and plot the power gain $G=20\log{|r^+|}$ as a function of $B_{\rm 0}$ in Fig.~\ref{fig:fig3}a. Away from resonance with the spins, we observe observe a dip in $G$ at $\omega_{\rm r}$, which is due to the internal resonator losses. As the spins come into resonance, they add energy to the system via stimulated emission, which compensates the internal losses in the resonator and transforms the dip in $G$ to a peak, signalling gain. We find a single gain peak, rather than the three that would be expected due to the hyperfine coupling of the NV$^-$ electronic spin with the nitrogen nuclear spin ($I=1$) \cite{barry2020sensitivity}. This is because of the inhomogeneity of the $B_{\rm 0}$ field supplied by the permanent magnet at the given pole separation, which broadens the spin transition and produces a Gaussian-like profile (see Supplementary Information, Section~\ref{sup:subsec:spectra}). 

We fix $B_{\rm 0}$ at the center of the gain feature in Fig.~\ref{fig:fig3}a (indicated by the dashed line) and study the effect of the optical power on the maser amplifier gain (Fig.~\ref{fig:fig3}b). At the lowest power applied $P_{\rm L} = 0.17$~W, the stimulated emission from the spins is not sufficient to overcome the cavity losses, however, as $P_{\rm L}$ is increased we observe rising levels of gain. By fitting the gain curves in Fig.~\ref{fig:fig3}b with our theoretical model (Eq.~\ref{eqn:r}), where we use independent measurements to tightly constrain some of the parameters (see Methods), we are able to extract the ensemble coupling strength $g_{\rm ens}$ as a function of $P_{\rm L}$ (inset Fig.~\ref{fig:fig3}b), which reaches a maximum of $g_{\rm ens}/2\pi = 1.54$~MHz at $P_{\rm L} = 1.61$~W.

We can write down an expression for the peak maser gain (which occurs at resonance $\omega = \omega_{\rm r}=\omega_{\rm s}$) using Eq.~\ref{eqn:r}:
\begin{align}\label{eqn:Gm}
    G_{\rm m} = |r^+|^2 = \frac{(\kappa_{\rm e}-\kappa_{\rm i}+\kappa_{\rm s})^2}{(\kappa_{\rm e}+\kappa_{\rm i}-\kappa_{\rm s})^2},
\end{align}
where we have defined a new parameter $\kappa_{\rm s} = 4g_{\rm ens}^2/\Gamma_{\rm eff}$ that describes the emission (absorption) rate of photons to (from) the resonator by the spin ensemble. The spin distribution effective width $\Gamma_{\rm eff}$ is in general a function of both $\Gamma$ and $\gamma$ (e.g. for a Lorentzian distribution $\Gamma_{\rm eff} = \Gamma + \gamma$) and is discussed in the Supplementary Information (Section~\ref{sup:sec:kc}). We note that this expression is identical to the one obtained by performing a conventional wave-approach analysis of a maser amplifier \cite{siegman1964microwave,mcwhorter1958solid}.

As $P_{\rm L}$ is increased we produce higher levels of spin polarization and larger values of $g_{\rm ens}$ (inset Fig.~\ref{fig:fig3}b), which raises the emission rate $\kappa_{\rm s}$. Inspecting Eq.~\ref{eqn:Gm} we can see that as $\kappa_{\rm s}$ (controlled by $P_{\rm L}$) approaches the total resonator loss rate $\kappa_{\rm e}+\kappa_{\rm i}$, the peak maser gain is expected to increase, as observed in our data (Fig.~\ref{fig:fig3}b). However, for $\kappa_{\rm s} = \kappa_{\rm e}+\kappa_{\rm i}$ the system reaches an instability point which defines the onset of maser oscillations \cite{jin2015proposal,breeze2018continuous}. This regime is undesirable for operating the system as an amplifier.

The maser is a negative resistance amplifier which exhibits a fixed gain-bandwidth-product (GBP) \cite{gordon1958noise}. We can trade peak gain for bandwidth by adjusting the external coupling rate $\kappa_{\rm e}$ (see Fig.~\ref{fig:fig3}c), which is controlled via the position of the loop coupler. We observe up to 30~dB of gain at small bandwidths and find the GBP varies between $1.7-4.5$~MHz over the range of $\kappa_{\rm e}$ explored here (see Fig.~\ref{fig:fig3}c inset).

\subsubsection{Compression power}\label{subsec:compression}
A key metric for an amplifier is its 1~dB compression point $P_{\rm 1dB}$, which describes the signal power at which the gain reduces by 1~dB, and is a useful measure of the linear operating range of the amplifier. We measure $P_{\rm 1dB}$ by sweeping the input microwave signal power $P_{\rm M}$ while monitoring the maser gain. Figure~\ref{fig:fig3}d shows the peak gain versus $P_{\rm M}$ for different laser powers $P_{\rm L}$. We find a constant compression power of $P_{\rm 1dB} = -73.9$~dBm (signal power at the output of the maser amplifier) over the range of gains explored. This corresponds to a projected -93.9~dBm of input power at 20~dB of gain, which is 20-30 dB higher than quantum-limited Josephson junction amplifiers \cite{zhou2014high,planat2019understanding,grebel2021flux}, but smaller than kinetic inductance based superconducting quantum-limited amplifiers \cite{parker2022degenerate} and about 10~dB below the recently demonstrated cryogenic NV maser amplifier~\cite{sherman2022diamond}. While the current performance is likely sufficient for applications that work with ultra-faint signals like deep space satellite communication \cite{macgregor2008low}, for others, including spin resonance spectroscopy, it would be beneficial to raise the $P_{\rm 1dB}$. In Section~\ref{sec:discussion} we discuss modifications to the system that could be made to push the compression power to higher levels.
 
\subsubsection{Noise temperature}\label{subsec:noise}
\begin{figure}[ht!]
\includegraphics[width
=\columnwidth]{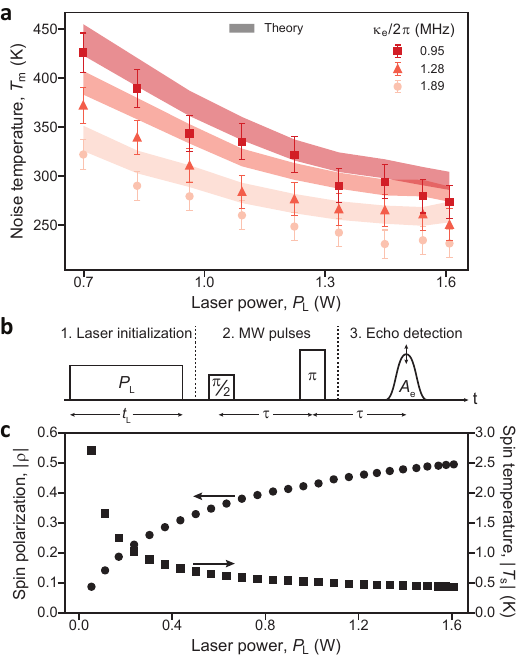}
\caption{\textbf{Maser amplifier noise temperature analysis.} \textbf{a}, Maser amplifier noise temperature as a function of the laser power, taken for different $\kappa_{\rm e}$. Measurements performed over a 1~kHz bandwidth at resonance ($\omega = \omega_{\rm r}=\omega_{\rm s}$). For the weakest coupling $\kappa_{\rm e}/2\pi=0.95$~MHz (strongest coupling $\kappa_{\rm e}/2\pi=1.89$~MHz), the peak maser gain is 7.3~dB (3.5~dB) at the smallest $P_{\rm L}$ and 17~dB (6.5~dB) at the highest $P_{\rm L}$. The shaded regions indicate the theoretically predicted range of noise temperatures based on the system parameters extracted from independent measurements. See Methods for a discussion on the error bars and model uncertainty. \textbf{b}, Schematic of experimental sequence used to determine the spin polarization $\rho$ and spin temperature $T_{\rm s}$ as a function of $P_{\rm L}$. In step 1 a laser pulse of duration $t_{\rm L} = 20$~ms and variable power $P_{\rm L}$ is used to initialize the spins. In step 2 a Hahn echo microwave (MW) pulse sequence, with delay time $\tau = 5~\mu$s, is delivered to the spins to induce a spin echo. In the final step the spins emit an echo, which is recorded and the amplitude $A_{\rm e}$ determined. \textbf{c}, Magnitudes of the spin temperature $|T_{\rm s}|$ and spin polarization $|\rho|$ (extracted from measurements of the spin echo amplitude) against laser power.}
\label{fig:fig4}
\end{figure}

We evaluate the maser amplifier's noise temperature $T_{\rm m}$ by utilizing a protocol based on the cold source measurement technique \cite{keysight2024high,pepe2017chip}. The setup (depicted in Fig.~\ref{fig:fig1}e) includes a spectrum analyzer (SA) or VNA connected to the maser system via a microwave circulator. A low-noise transistor amplifier boosts the output of the maser before detection. Through precise measurement of the transistor amplifier's noise temperature, the insertion loss of the system components, as well as the maser amplifier's output noise power and gain, we are able to extract the intrinsic noise temperature of the maser system (see Methods and Supplementary Information, Section~\ref{sup:sec:noise}).

Figure~\ref{fig:fig4}a shows the extracted maser amplifier noise temperature (referred to its input) versus laser power $P_{\rm L}$ for different values of the external coupling rate $\kappa_{\rm e}$. As $P_{\rm L}$ is increased, all datasets demonstrate a reduction in the noise temperature. In addition, we observe that increasing $\kappa_{\rm e}$ also reduces the noise temperature. We find a minimum noise temperature of $T_{\rm m} = 231(14)$~K for $\kappa_{\rm e} = 1.89$~MHz, where the maser gain is 6.5~dB. To understand this behavior, we consider our theoretical expression for the output noise (Eq.~\ref{eqn:nout}). The output noise $n_{\rm out}^+$ contains three contributions, amplified noise from the input field, amplified noise from the spins and amplified noise from the resonator losses. The first component is not intrinsic to the amplifier and is subtracted in our measurements, whilst the remaining two contributions dictate the maser noise temperature.

To provide an intuition as to how the various parameters affect the noise, we define a new quantity $n_{\rm m} = n_{\rm out}^+/R_{\rm in}^+-\left(n_{\rm in} + 1/2\right)$ which is the input-referred maser amplifier noise (i.e. the maser's output noise divided by its gain). At resonance ($\omega = \omega_{\rm r}=\omega_{\rm s}$) $n_{\rm m}$ becomes:
\begin{align}
    n_{\rm m} &= \frac{G_{\rm s}}{G_{\rm m}}\left(n_{\rm s} + \frac{1}{2}\right) + \frac{G_{\rm i}}{G_{\rm m}}\left(n_{\rm i} + \frac{1}{2}\right),\label{eqn:nm}
\end{align}
where,
\begin{equation}
\begin{aligned}\label{eqn:GsGi}
    \frac{G_{\rm s}}{G_{\rm m}} &= \frac{4\kappa_{\rm e}\kappa_{\rm s}}{\left(\kappa_{\rm e}-\kappa_{\rm i}+\kappa_{\rm s}\right)^2},\\
    \frac{G_{\rm i}}{G_{\rm m}} &= \frac{4\kappa_{\rm e}\kappa_{\rm i}}{\left(\kappa_{\rm e}-\kappa_{\rm i}+\kappa_{\rm s}\right)^2}.
\end{aligned}
\end{equation}

Once again, we find that the simplified expressions in Eqs.~\ref{eqn:nm} and \ref{eqn:GsGi} agree exactly with those obtained from a simple wave-approach analysis of a negative resistance amplifier \cite{siegman1964microwave,mcwhorter1958solid}. It is evident from Eq.~\ref{eqn:GsGi} that both components of $n_{\rm m}$ reduce with increasing $\kappa_{\rm s}$ (and hence $P_{\rm L}$) and also with increasing $\kappa_{\rm e}$, as seen in our measurements. We note, however, that whilst increasing $\kappa_{\rm s}$ enhances the gain and reduces noise, increasing $\kappa_{\rm e}$ (at a fixed $\kappa_{\rm s}$) reduces the noise at the expense of lowering the gain (see Eq.~\ref{eqn:Gm}).

To make theoretical predictions of $n_{\rm m}$, we extract the parameters $\kappa_{\rm e}$, $\kappa_{\rm i}$, $g_{\rm ens}$ and $\Gamma_{\rm eff}$ at each data point in Fig.~\ref{fig:fig4}a from independent measurements (see Methods) and assume that the resonator bath is at the physical temperature of the resonator ($T_{\rm i} \approx 294$~K). We estimate the spin bath occupation $n_{\rm s}$ from measurements of the effective spin temperature, as detailed below. We then convert $n_{\rm m}$ to an effective temperature by inverting the Bose-Einstein relation $T_{\rm m} = \hbar\omega_{\rm s}/[k_{\rm B}\ln{\left(1+1/n_{\rm m}\right)}]$, allowing us to predict the range of noise temperatures expected in Fig.~\ref{fig:fig4}a (shaded regions) from uncertainties in the model parameters. We find good agreement overall, with all theoretical predictions residing within 15\% of the measured temperatures. 

To examine the limits of noise in a maser amplifier, we take the extreme case of a lossless resonator ($\kappa_{\rm i} = 0$) with a spin bath at zero temperature ($n_{\rm s} = 0$). We see that for large gain ($\kappa_{\rm s} \rightarrow \kappa_{\rm e}$) the maser noise approaches the quantum limit for a phase-preserving amplifier \cite{caves1982quantum}, i.e. $n_{\rm m} \rightarrow 1/2$, corresponding to a temperature $T_{\rm m} = \hbar\omega_{\rm s}/[k_{\rm B}\ln{\left(3\right)}]\approx\hbar\omega_{\rm s}/k_{\rm B}$. However, for imperfect optical pumping, the effective spin temperature is non-zero. The spin temperature is defined by the spin polarization $\rho$, which is given as:
\begin{equation}
\rho = \tanh \left(\frac{\hbar\omega_{\rm s}}{2k_{\rm B}T_{\rm s}}\right).\label{eqn:rhotanh}
\end{equation}

The spin polarization in our system is determined by optical pumping, where $\rho = (N_{-1}-N_{0})/(N_{-1}+N_{0})$, with $N_{-1}$ ($N_{0}$) representing the population of the $m_{\rm s}=-1$ ($m_{\rm s}=0$) spin state. Since $N_{0}>N_{-1}$, the spin temperature at the maser amplification transition is negative. 

We determine $\rho$ (and thus $T_{\rm s}$) in our device by performing Hahn echo measurements using the protocol depicted in Fig.~\ref{fig:fig4}b. The sequence starts with a spin initialization phase, where the diamond is exposed to light at a power $P_{\rm L}$ for a duration of $t_{\rm L} = 20$~ms. Following this, a Hahn echo pulse sequence is delivered to the spins and the resulting spin echo signal is recorded in a final step. We find the amplitude of the spin echo $A_{\rm e}$, which is proportional to $\rho$ \cite{bienfait2016reaching}. When $P_{\rm L} = 0$~W (no optical initialization), the the spins are in thermal equilibrium at room temperature ($T_0\approx 294$~K) and $\rho$ can be determined from Eq.~\ref{eqn:rhotanh} by setting $T_{\rm s} = T_0$. Taking this absolute value of $\rho$, together with the ratio of echo amplitudes measured with the laser on versus the laser off, we are able to calculate $\rho$ and $T_{\rm s}$ as a function of $P_{\rm L}$ (see Supplementary Information, Section~\ref{sup:sec:spin_temp}). We plot the resulting magnitudes of the spin polarization $|\rho|$ and the spin temperature $|T_{\rm s}|$ in Fig.~\ref{fig:fig4}c.

For the largest $P_{\rm L}$ applied, we achieve a maximum spin polarization of $|\rho| \approx 0.5$, corresponding to an effective spin temperature $|T_{\rm s}| < 0.5$~K. We therefore estimate the intrinsic noise contribution from the spin gain medium (first term in Eq.~\ref{eqn:nm}) to be between $1-2$~K for the range of $\kappa_{\rm e}$ explored in Fig.~\ref{fig:fig4}a at maximum laser power. The maser amplifier noise temperature is therefore almost entirely determined by resonator losses. In Section~\ref{sec:discussion} we discuss viable pathways to reduce the noise originating from the resonator loss thermal bath.

\subsection{Microwave cooling}\label{sec:cooling}
\begin{figure*}[ht!]
\includegraphics[width=2\columnwidth]{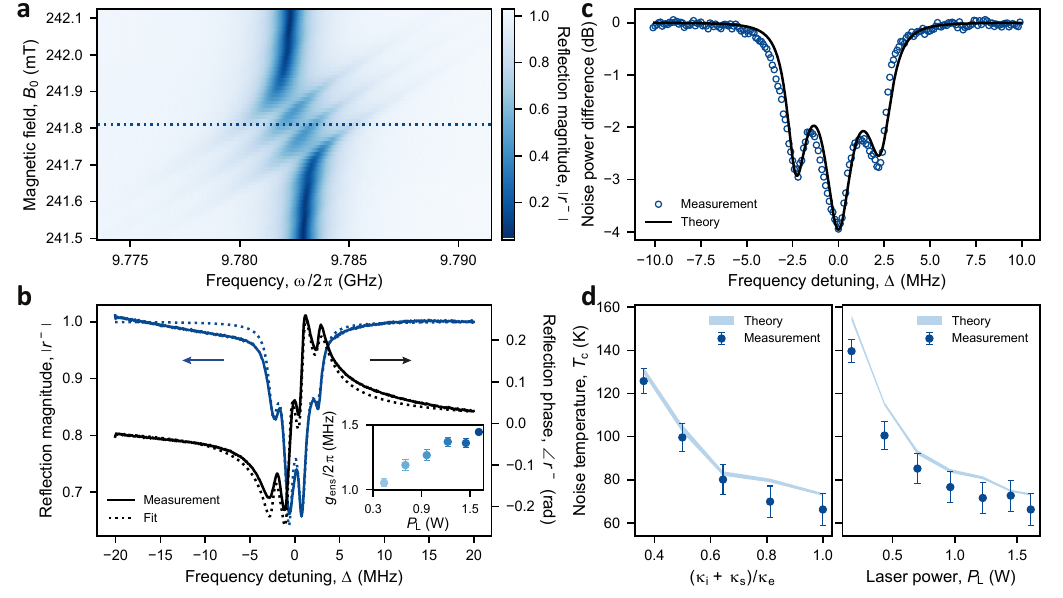}
\caption{\textbf{Microwave cooling measurements.} \textbf{a}, Reflection magnitude response $|r^-|$ as $B_{\rm 0}$ is swept across the cooling NV$^-$ spin transition. The dotted line represents a line-cut at which the measurement in panel b is taken. \textbf{b}, Line-cut of the reflection coefficient magnitude $|r^-|$ (blue) and phase $\angle r^-$ (black) responses, both fit to input-output theory. Inset shows the $g_{\rm ens}$ extracted from the fits as a function of $P_{\rm L}$. \textbf{c}, Reduction in the noise power with the spins on resonance and the system critically coupled ($\kappa_{\rm e} = \kappa_{\rm i} + \kappa_{\rm s}$). The open circles are measured data whilst the solid black line is the prediction from input-output theory with no free fitting parameters. \textbf{d}, Left: Microwave cooler noise temperature as a function of the coupling ratio $(\kappa_{\rm i}+\kappa_{\rm s})/\kappa_{\rm e}$, taken at maximum laser power. Right: Microwave cooler noise temperature as a function of the laser power, taken at critical coupling ($\kappa_{\rm e} = \kappa_{\rm i} + \kappa_{\rm s}$).  Measurements performed over a 10~kHz bandwidth at resonance ($\omega = \omega_{\rm r}=\omega_{\rm s}$). The blue shaded regions are the range of cooler noise temperatures predicted with input-output theory. See Methods for details regarding the theoretical predictions and a discussion on the error bars.}
\label{fig:fig5}
\end{figure*}

Next we adjust the permanent magnet pole positions to set $B_{\rm 0} \approx 250$~mT, bringing the $m_{\rm s} = 0$ to $m_{\rm s} = +1$ spin transition in resonance with the dielectric resonator. The NV$^-$ spins, which are now polarized in the ground state of this two-level subsystem, absorb microwave photons from the resonator (including thermal photons), which are then removed from the system via optical pumping. The removal of thermal noise photons lowers the effective microwave temperature of the mode \cite{wu2021bench,ng2021quasi,zhang2022microwave,fahey2023steady,blank2023anti}, which can be transferred to an external microwave circuit. 

The noise emitted into the circuit by the microwave cooler $n_{\rm c}$ can be found using Eq.~\ref{eqn:nout}, i.e. $n_{\rm c} = n_{\rm out}^-$. At resonance ($\omega = \omega_{\rm r}=\omega_{\rm s}$), we have:
\begin{equation}\label{eqn:Rminus}
\begin{aligned}
    R_{\rm in}^- &= \frac{\left(\kappa_{\rm e}-\kappa_{\rm i}-\kappa_{\rm s}\right)^2}{\left(\kappa_{\rm e}+\kappa_{\rm i}+\kappa_{\rm s}\right)^2},\\
    R_{\rm s}^- &= \frac{4\kappa_{\rm e}\kappa_{\rm s}}{\left(\kappa_{\rm e}+\kappa_{\rm i}+\kappa_{\rm s}\right)^2},\\
    R_{\rm i}^- &= \frac{4\kappa_{\rm e}\kappa_{\rm i}}{\left(\kappa_{\rm e}+\kappa_{\rm i}+\kappa_{\rm s}\right)^2},
\end{aligned}
\end{equation}
which determine the contributions to the cooler output noise from the input field, spin bath, and resonator losses, respectively. At resonance the cooler noise is minimized when the system is critically coupled $\kappa_{\rm e}=\kappa_{\rm i}+\kappa_{\rm s}$. Under these conditions $n_{\rm c}$ simplifies to:
\begin{align}\label{eqn:ncool}
    n_{\rm c} &= \frac{\kappa_{\rm s}}{\kappa_{\rm e}}\left(n_{\rm s} + \frac{1}{2}\right) + \frac{\kappa_{\rm i}}{\kappa_{\rm e}}\left(n_{\rm i} + \frac{1}{2}\right),
\end{align}
which reduces as $\kappa_{\rm e}$ and $\kappa_{\rm s}$ grow relative to $\kappa_{\rm i}$.

The system acts like a perfectly matched and cryogenically-cooled $50~\Omega$ termination, which absorbs all of the room-temperature microwave noise from the input field (i.e., $R_{\rm in}^-=0$) and emits cold noise at a temperature $T_{\rm c} = \hbar\omega_{\rm s}/[k_{\rm B}\ln{\left(1+1/n_{\rm c}\right)}]$ into the circuit. 

To find the cooling spin transition we perform a fine sweep of $B_{\rm 0}$ whilst probing the resonator in reflection, with the result shown in Fig.~\ref{fig:fig5}a. We observe avoided crossings in the measured spectrum, which is indicative of strong coupling between the resonator and hyperfine transitions of the NV$^-$ spins \cite{zhang2022cavity,fahey2023steady}. We note that at this magnet pole separation we are able to partially resolve the hyperfine structure of the NV$^-$ system, indicating a higher field homogeneity than observed at the maser transition. Taking a line cut (dashed line in Fig.~\ref{fig:fig5}a) through the middle of this feature, we observe four dips in the magnitude response (see Fig.~\ref{fig:fig5}b, dark blue line), which are reproduced well with our model (see Eq.~\ref{eqn:r}). A fit to the magnitude ($|r^-|$) and phase ($\angle r^-$) reflection data allows us to extract the ensemble coupling strength $g_{\rm ens}$, which is plot as an inset in Fig.~\ref{fig:fig5}b as a function of $P_{\rm L}$. We find excellent agreement with the $g_{\rm ens}$ extracted via the gain measurements in Section~\ref{subsec:gain}.

To demonstrate cooling, we monitor the system noise using a SA with the setup depicted in Fig.~\ref{fig:fig1}e. We use a room-temperature low-noise high electron mobility transistor (HEMT) amplifier with a noise temperature of $T_{\rm A}=47(10)$~K (see Methods) to boost the system output above the noise floor of the SA. We fix $B_{\rm 0}$ at the center of the spin resonance feature and change the position of the loop coupler until the system is at critical coupling $\kappa_{\rm e}=\kappa_{\rm i}+\kappa_{\rm s}$. In Fig.~\ref{fig:fig5}c we plot the the difference (in decibels) between two noise power measurements, (i) with the spins on resonance and the laser on, and (ii) with the spins off resonance (i.e., $B_{\rm 0}$ detuned) and the laser off. We see that for frequencies within the spin line width, the presence of the spins lowers the system noise by up to $\sim 4$~dB (i.e., a factor of $2.5\times$), corresponding to a system noise temperature of $(T_{\rm A}+T_0)/2.5 \approx 136$~K at resonance ($\omega=\omega_{\rm s}$). Further, we find an average 2.8~dB of noise reduction over 5~MHz of bandwidth centered on $\omega_{\rm s}$. We predict the cooling performance of the system using Eq.~\ref{eqn:nout}, where we enter all parameters (extracted from independent measurements, see Methods) into the model and plot the result in Fig.~\ref{fig:fig5}c (solid black line). The model quantitatively reproduces the data without any free parameters.

We can calculate the cooler noise temperature $T_{\rm c}$ by removing from the system noise any contribution due to insertion loss of the cables and components as well as the noise added by the HEMT amplifier (see Supplementary Information, Section~\ref{sup:sec:cooling}). In Fig.~\ref{fig:fig5}d we plot the inferred $T_{\rm c}$ as a function of the coupling ratio $(\kappa_{\rm i}+\kappa_{\rm s})/\kappa_{\rm e}$ (left panel). As expected, the temperature reaches its lowest value at critical coupling $(\kappa_{\rm i}+\kappa_{\rm s})/\kappa_{\rm e}=1$. In the right panel of Fig.~\ref{fig:fig5}d we plot the cooler noise temperature versus laser power, ensuring that the system is critically coupled at each point. We find that $T_{\rm c}$ reaches a minimum of $66(7)$~K. Once again, our theory (light blue bands in Fig.~\ref{fig:fig5}d) is able to closely predict the noise performance of the microwave cooler for both experiments.

\section{Discussion}\label{sec:discussion}
Our experiments have demonstrated a continuous-wave solid-state maser amplifier operating at room temperature. The system is made from simple components (such as a laser diode and permanent magnet) which could be translated to a small and portable form factor. Simply by adjusting the magnetic field we were able to switch to a mode where the system behaved as a cryogenically-cooled matched load, which emitted noise at a temperature of $66(7)$~K into a microwave circuit. The measured noise temperature of our maser amplifier is comparable to current commercial general-purpose low-noise microwave amplifiers, but about a factor of 5 worse than state-of-the-art HEMT amplifiers. In addition, the GBP of the maser (up to $\sim 4.5$~MHz) currently only permits narrow band operation, with $\sim 10$~dB of gain over a $\sim 0.8$~MHz bandwidth. Whilst the current level of performance represents a promising first step, there are several avenues available to push its operation towards the quantum noise limit, extend the available bandwidth and improve its power handling capabilities. 

The intrinsic noise generated by the spin gain medium ($\sim 1-2$~K) was found to already be close to the quantum noise limit ($\sim 0.5$~K), thus the main route to improve the noise performance is to reduce the contribution due to resonator losses (second term, Eq.~\ref{eqn:nm}). This can be achieved by decreasing $\kappa_{\rm i}$ and/or increasing $\kappa_{\rm s}$ -- the latter also simultaneously improving the GBP. Alternative ceramics or low-loss dielectric crystals could be explored to produce resonators with higher internal quality factors $Q_{\rm i}$. Increasing the concentration of NV$^-$ spins and/or the filling factor of the diamond sample inside the mode's magnetic field would enhance $g_{\rm ens}$ and therefore $\kappa_{\rm s}$ (see Supplementary Information, Section~\ref{sup:subsec:DR}). We estimate that doubling the NV$^-$ concentration from the current 3~ppm to 6~ppm and raising the internal quality factor from 10,000 to 20,000 would be sufficient to double the bandwidth and push the noise temperature below 30~K, surpassing the noise performance of the best commercially available room-temperature microwave X-band amplifiers. Such improvements are within reach. Increasing the concentration and therefore number of NV$^-$ spins would also directly enhance the compression power $P_{\rm 1dB}$ \cite{jin2015proposal}.

Looking further, we could improve the filling factor from the current 11\% to 35\% (i.e., a factor of $\sim 3$) by filling the entire center of the cylindrical dielectric resonator with NV diamond. We could also move from CVD to high pressure and high temperature (HPHT) grown diamonds, where higher NV$^-$ concentrations of up to 20~ppm are possible \cite{blank2023anti}. This could boost the GBP and $P_{\rm 1dB}$ by a factor of $20\times$ relative to the present demonstration, and together with a doubling of $Q_{\rm i}$, lower the noise temperature to 4~K. One could utilize even lower loss sapphire dielectric resonators, with $Q_{\rm i} > 50,000$ \cite{breeze2018continuous}, pushing the noise temperature towards 1~K. However, the lower dielectric constant of sapphire ($\epsilon_{\rm r}\approx 9.4$) would increase the mode volume and require even larger diamond samples to maintain high filling factors, leading to increased laser powers and stricter requirements for thermal management.

The GBP could be enhanced even further by moving to travelling wave \cite{degrasse1959three} or reflected wave \cite{moore1979reflected} architectures, which eliminate resonators to increase the bandwidth. However, challenges in producing large volume high-concentration NV diamond samples might mean taking a hybrid approach between reflected wave and resonator designs, such as a multi-stage resonant amplifier \cite{shell1994dual}, is more feasible.

The changes noted above would also serve to improve the achievable levels of microwave cooling. In the present setup, insertion loss ($\sim 0.52$~dB between the device and the HEMT amplifier) limits the largest amount of noise reduction possible to 6.4~dB. This could be improved by transitioning from relatively lossy coaxial cables to waveguide technology \cite{zollitsch2023maser}. By combining both maser amplification and microwave cooling in one experiment, one could ensure all components of the system noise are reduced, permitting the room-temperature detection of microwave signals with cryogenic-levels of noise.

This technology could one day replace cryogenically-cooled microwave receivers used for satellite communication in the deep space network \cite{macgregor2008low}, radio astronomy \cite{wilson2009tools}, or be utilized to permit high-sensitivity spin resonance spectroscopy at room temperature. These results show that NV spin ensembles in diamond strongly coupled to high quality factor microwave resonators form an exceptional system for performing low noise microwave measurements under ambient conditions.

\section{Methods}\label{sec:Methods}
\subsection{Diamond sample}
The NV diamond sample was custom manufactured by HiQuTe Diamond. It was grown via chemical vapor deposition (CVD) with $\{100\}$ faces and doped with a high concentration of nitrogen using N$_2$O as the precursor. The sample was then laser cut and polished to a rectangular prism of dimensions 1.8~mm~$\times~$1.9~mm~$\times$~2.0~mm before being irradiated with 3~MeV electrons at a fluence of $1\times10^{18}$~cm$^{-2}$ and then annealed. Following this the sample underwent acid (aqua regia, 1 hour at $100^\circ$C) and hydrogen plasma cleaning to remove any residual traces of graphite. The resulting concentration of NV$^-$ was estimated to be 3~ppm by performing UV-visible absorption spectroscopy at a temperature of 12~K on a thinner sample that was grown and processed under the same conditions. 

\subsection{Permanent magnet system}\label{sec:Methods:magnet}
A compact, commercial permanent magnet system (Spinflex Instruments Ltd.) was used to provide the static $B_{\rm 0}$ field for tuning the NV$^-$ spin transitions into resonance with the fixed frequency dielectric resonator. The magnet features a small Helmholtz coil for fine-tuning the field over a range of $\sim 50$~mT (for currents up to 1~A) around the permanent magnet operating point. The coils are controlled via a closed loop feedback system that measures the magnetic field 18~mm from the center of the sample with a Hall probe and adjusts the current in the coils accordingly to keep the measured magnetic field at the set point. Larger shifts in the $B_{\rm 0}$ operating point were achieved by changing the separation between the two permanent magnet poles, with the lower field ($\sim242$~mT) cooling transition requiring a pole separation of 38~mm and the higher field ($\sim450$~mT) maser amplification transition requiring a pole separation of 23~mm. The permanent magnet was designed to have an optimal field homogeneity for pole separations between 40~mm and 50~mm, achieving a nominal homogeneity of approximately $10~\mu$T over a 5~mm diameter sphere. In our pulsed ESR characterization of the sample (see Supplementary Information, Section \ref{sup:subsec:spectra}) we observe an increased broadening of the NV$^-$ spin transition as we decrease the pole spacing from the cooling to the maser amplification operating point. We attribute this to inhomogenous broadening caused by a decrease in the magnetic field homogeneity over the sample as we move further away from the optimal pole separation.

\subsection{Microwave measurement setup}
Microwave measurements were performed by probing the resonator via an inductive loop coupler attached to a linear piezoelectric stage, allowing the external coupling $\kappa_{\rm e}$ to be varied by changing the distance between the loop and the resonator. Most measurements were completed using the circuitry depicted in Fig.\ref{fig:fig1}e, where a dual-use vector network and spectrum analyser (Keysight FieldFox, model N9918B) was connected to the coupler via a circulator.

The device gain was found from vector network analyzer (VNA) reflection measurements, where a low power probe tone was directed through the circulator to the device, and the reflected amplified signal was sent through a low-noise transistor amplifier before being recorded by the analyzer. When off resonance with the mode frequency, the probe tone reflects from the loop coupler short circuit with $|r^{\pm}| = 1$. The gain of the maser amplifier can therefore be found by normalizing the traces with the off resonance reflection magnitude.

For the noise temperature and cooling measurements, we switched the FieldFox to its spectrum analyzer (SA) mode. Here the probe tone is turned off and the device sees noise coming from an effective 50~$\Omega$ room-temperature ($294$~K) load.

\subsection{Gain measurements}\label{sec:Methods:gain}
When fitting the gain curves in Fig.~\ref{fig:fig3}b with our input-output theory (Eq.~\ref{eqn:r}) to extract $g_{\rm ens}$, we make use of independently measured parameters to constrain the fit. Parameters describing the resonator $\kappa_{\rm e}$, $\kappa_{\rm i}$ and $\omega_{\rm r}$ are extracted from reflection measurements using a `circle fit' algorithm \cite{probst2015efficient}. The $K(\omega)$ function is constrained by the spin linewidth $\Gamma_{\rm eff}$, obtained through fitting the maser amplification transition ESR spectra with a Gaussian function (see Supplementary Information, Section~\ref{sup:subsec:spectra}). The coupling strength $g_{\rm ens}$ is the only remaining parameter and was allowed to vary over a broad range.
 
\subsection{Noise temperature measurements}\label{sec:Methods:noise}
We use a commercial VNA (Keysight, PNA-X) to measure noise temperatures using the cold-source technique \cite{keysight2024high, pepe2017chip}. This technique combines accurate gain measurements with noise power measurements using a precisely calibrated analyzer to acquire the noise temperature with a high level of accuracy. The PNA-X does not facilitate accurate noise temperature measurements for bandwidths below 800~kHz. We therefore use the PNA-X to measure the noise temperature of the wide band transistor amplifier (Minicircuits ZVA-183-S+) and then combine this with an accurate calibration of the insertion loss of the cables and circulator to find the maser noise temperature, using the procedure detailed in the Supplementary Information (Section~\ref{sup:sec:noise}). Error bars on the data points in Fig.~\ref{fig:fig4}a are found by combining the uncertainty (one standard deviation) on the gain, noise power and second amplifier noise temperature measurements with our estimate for the uncertainty in the insertion loss ($\pm 0.1$~dB).

To predict the maser amplifier noise temperature with our input-output theory (Fig.~\ref{fig:fig4}), we require knowledge of $\kappa_{\rm e}$, $\kappa_{\rm i}$, $\omega_{\rm r}$, $\omega_{\rm s}$, $g_{\rm ens}$,  and $\Gamma_{\rm eff}$ at each operating point. The coupling parameters $\kappa_{\rm e}$ and $\kappa_{\rm i}$, along with the mode frequency $\omega_{\rm r}$ and thus $\omega_{\rm s}$ (since $\omega_{\rm s}=\omega_{\rm r}$) are extracted from reflection measurements using a `circle fit' algorithm \cite{probst2015efficient}. The reflection measurements were taken immediately before the series of noise temperature measurements, with $B_{\rm 0}$ and therefore the spin transition detuned. The coupling strength $g_{\rm ens}$ was extracted from gain measurements and the spin linewidth $\Gamma_{\rm eff}$ from pulsed ESR spectroscopy measurements, in the same manner as in Section~\ref{sec:Methods:gain}. We calculated the range of model predictions in Fig.~\ref{fig:fig4}a by using the uncertainties (one standard deviation) of the parameters to find the extreme values of the noise temperature.

\subsection{Cooling measurements}
For the cooling measurements we use a low-noise HEMT amplifier (Low Noise Factory LNF-LNC0.3\_14A), with a noise temperature of $\sim 44$~K at 10~GHz and 296~K (as per the manufacturer datasheet). The HEMT amplifier was connected directly to the output port of the low-loss circulator to boost the system noise above the noise floor of the SA. We measured the noise temperature of the HEMT amplifier using the cold source method \cite{keysight2024high}, finding a value of $47(10)$~K, consistent with the manufacturer specifications. Error bars on the data points in Fig.~\ref{fig:fig5}d are found by combining the uncertainty (one standard deviation) on the noise power and second amplifier noise temperature measurements with our estimate for the uncertainty in the insertion loss ($\pm 0.1$~dB).

The parameters used to predict the frequency dependent cooling profile in Fig.~\ref{fig:fig5}c were found from independent reflection and ESR measurements. The spin linewidths were extracted by fitting the hyperfine-split ESR spectrum to Lorentzian functions (see Supplementary Information, Section~\ref{sup:subsec:spectra}). The rates $\kappa_{\rm e}$ and $\kappa_{\rm i}$ and frequency $\omega_{\rm r}$ were found from reflection measurements taken with the spins detuned from the resonator frequency. We determine $g_{\rm ens}$ by fitting a reflection measurement with our input-output theory (Eq.~\ref{eqn:r}), as shown in Fig.~\ref{fig:fig5}b. In Fig.~\ref{fig:fig5}d (left panel) we use Eq.~\ref{eqn:nout} and the coefficients listed in Eq.~\ref{eqn:Rminus} (which are valid at resonance) to make predictions of the microwave cooler output noise temperature. The rates $\kappa_{\rm e}$ and $\kappa_{\rm i}$ are extracted from resonator reflection measurements recorded with the spins detuned from resonance, whilst $\kappa_{\rm s}$ is found from a reflection measurement performed with the spins at resonance. We calculate a range in model predictions from the uncertainties (standard deviations) of the model parameters. The model predictions in Fig.~\ref{fig:fig5}d (right panel) are found from the simplified cooler noise expression in Eq.~\ref{eqn:ncool} and calculated in the same manner as the left panel.

\bibliographystyle{unsrt}
\bibliography{bib}
\vspace{1cm}

\noindent{\small{\textbf{Acknowledgments:} We acknowledge support from the NSW Node of the Australian National Fabrication Facility. This work used the NCRIS and Government of South Australia enabled Australian National Fabrication Facility - South Australian Node (ANFF-SA). We acknowledge Evan Johnson and Luis Lima-Marques from the Optofab Adelaide Node of the Australian National Fabrication Facility for fabricating the ceramic dielectric resonators. We thank Klaus Mølmer, Yuan Zhang and QiLong Wu for helpful discussions on the input-output theory developed in this work. We thank Riadh Issaoui from Hiqute Diamond for producing the high-concentration NV diamond sample used in this study. We acknowledge Alex Rizgalla from Keysight for providing the PNA-X vector network analyzer used in the noise temperature measurements. Finally, we thank Tony Melov for producing the artist's impression of the NV diamond amplifier/cooler setup in Fig.~\ref{fig:fig1}D and Fig.~\ref{fig:fig2}. \textbf{Funding:} J.J.P. acknowledges support from an Australian Research Council Future Fellowship (FT220100599). T.D. and M.I. acknowledge financial support from Sydney Quantum Academy, Sydney, NSW, Australia. This research has been supported by an Australian Government Research Training Program (RTP) Scholarship. \textbf{Author contributions:} J.J.P. and T.D. designed the experiment. M.I. assisted with the design of the optics setup. T.D. performed the measurements and T.D. and J.J.P. analyzed the data. M.I. and W.J.P. helped with sample characterization. B.C.J., H.A., and T.O. provided materials and assisted with processing diamond samples. J.J.P. developed the input-output theory. J.J.P., A.L., and D.R.M. supervised the project. T.D. and J.J.P. wrote the manuscript with input from all authors. \textbf{Competing interests:} The authors declare that they have no competing interests. \textbf{Data and materials availability:} All data needed to evaluate the key conclusions in the paper are present in the paper and/or the Supplementary Information. Additional data is available from the corresponding author upon reasonable request. \textbf{Additional information:} Online Supplementary Information accompanies this paper. Correspondence should be addressed to J.J.P.}}

\end{document}


\title{Supplementary Information: A Room-Temperature Solid-State Maser Amplifier}

\author{Tom Day}
\affiliation{School of Electrical Engineering and Telecommunications, UNSW Sydney, Sydney, NSW 2052, Australia}
\author{Maya Isarov}
\affiliation{School of Electrical Engineering and Telecommunications, UNSW Sydney, Sydney, NSW 2052, Australia}
\author{Billy Pappas}
\affiliation{School of Physics, UNSW Sydney, Sydney, NSW 2052, Australia}
\author{Brett C. Johnson}
\affiliation{School of Science, RMIT University, Melbourne, Victoria 3001, Australia}
\author{Hiroshi Abe}
\affiliation{National Institutes for Quantum Science and Technology (QST), 1233 Watanuki, Takasaki, Gunma 370-1292, Japan}
\author{Takeshi Ohshima}
\affiliation{National Institutes for Quantum Science and Technology (QST), 1233 Watanuki, Takasaki, Gunma 370-1292, Japan}
\affiliation{Department of Materials Science, Tohoku University, Aoba, Sendai, Miyagi 980-8579, Japan}
\author{Dane McCamey}
\affiliation{School of Physics, UNSW Sydney, Sydney, NSW 2052, Australia}
\author{Arne Laucht}
\affiliation{School of Electrical Engineering and Telecommunications, UNSW Sydney, Sydney, NSW 2052, Australia}
\author{Jarryd J. Pla}
\email[]{jarryd@unsw.edu.au}
\affiliation{School of Electrical Engineering and Telecommunications, UNSW Sydney, Sydney, NSW 2052, Australia}
\date{\today}

\maketitle

\tableofcontents

\newcommand{\beginsupplement}{%
\setcounter{table}{0}
\renewcommand{\thetable}{S\arabic{table}}%
\setcounter{figure}{0}
\renewcommand{\thefigure}{S\arabic{figure}}%
}
\beginsupplement

\section{Experimental setup}\label{sup:sec:setup}
\subsection{Dielectric resonator}\label{sup:subsec:DR}
\begin{figure}[h]
\includegraphics{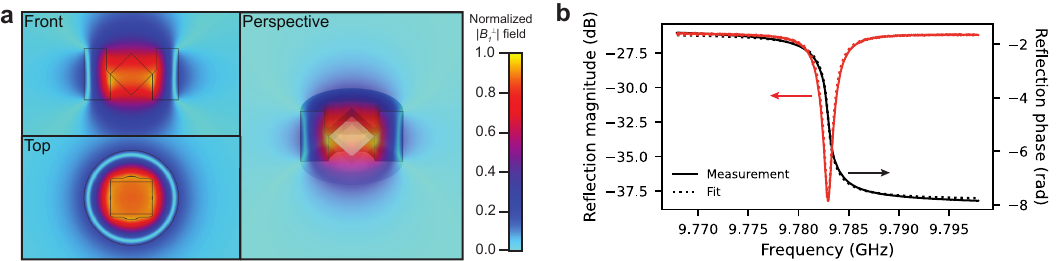}
\caption{\textbf{Dielectric resonator simulation and measurement.} \textbf{a} CST Studio Suite mode profile simulation showing the magnitude of the perpendicular component of the magnetic field across different views (front, top and perspective). \textbf{b}, Reflection coefficient magnitude (red) and phase (black) measurement (solid line) of the resonator loaded with the diamond sample, overlaid with a fit (dotted line) used to extract the coupling rates.}
\label{sup:fig:DR}
\end{figure}

The dielectric resonator material choice and design is key to extracting the most performance out of the maser amplifier/cooler setup. The magnetic field profile of the resonator mode sets the coupling strength $g_{\rm ens}$ to the NV$^-$ spin ensemble and the microwave loss tangent of the dielectric material dictates the rate at which room-temperature noise is mixed into the system (see Section~\ref{sup:sec:Theory}). We define the volume of the mode's magnetic field (or the `magnetic mode volume') as:
\begin{equation}
V_{\rm m} = \frac{\iiint_\text{total}\left|B_1\right|^2}{\max\left(\left|B_1\right|^2\right)}
\end{equation}
where $\left|B_1\right|^2 = \left|B_1^{\rm x}\right|^2 + \left|B_1^{\rm y}\right|^2 + \left|B_1^{\rm z}\right|^2$ is the squared magnitude of the magnetic field, with components $(B_1^{\rm x},B_1^{\rm y},B_1^{\rm z})$ at a particular point in space. The numerator is the volume integral of the squared field over all space whilst the denominator is the maximum value of $\left|B_1\right|^2$ taken over the entire mode. In our experiment the sample volume is $V_{\rm s}=6.84$~mm$^3$, whilst $V_{\rm m} = 45.8$~mm$^3$, as determined from finite-element simulations (see Fig.~\ref{sup:fig:DR}a and discussion below). In this situation where the sample does not occupy all of the mode's magnetic field, the energy of the magnetic field that resides within the sample becomes an important figure of merit. This is known as the magnetic filling factor:
\begin{equation}
\eta = \frac{\iiint_{\text{sample}}\left|B_1^\perp\right|^2}{\iiint_{\text{total}}\left|B_1\right|^2}
\end{equation}
where $|B_1^\perp| = \sqrt{|B_1^{\rm y}|^2 + |B_1^{\rm z}|^2}$ is the relevant component of the field that couples to the spins, assuming one of the NV $\left<111\right>$ axes and the $B_0$ field are aligned along the x-axis.

The spin ensemble coupling strength scales with the filling factor as \cite{grezes2016towards} $g_{\rm ens} \propto \sqrt{\eta}$, so it is desirable to make $\eta$ as high as possible. This can achieved by shrinking the magnetic mode volume down to the sample volume $V_{\rm s}$. The magnetic mode volume can be reduced by using materials with high dielectric constants $\epsilon_{\rm r}$, however, finding materials with high $\epsilon_{\rm r}$ and low microwave losses can be challenging. We utilize a low loss ceramic (MuRata Manufacturing Co., Ltd.), which has a nominal internal quality factor of $Q_{\rm i} = 12,000$ at 10~GHz while maintaining a relatively high dielectric constant ($\epsilon_{\rm r} \approx 30$). Additionally, the material has a small resonant frequency temperature coefficient (nominally 0~ppm/$^{\circ}$K), limiting the effect of any incident laser power or changes in the ambient temperature on the resonators stability.

We designed our resonator geometry using CST Studio Suite (see Figure \ref{sup:fig:DR}a) to ensure optimal performance within our material, sample size and manufacturing tolerance constraints. We achieve a filling factor of $\eta = 0.11$, which could be increased as high as $\eta = 0.35$ for the current design by filling the entire central bore of the dielectric resonator with diamond. A measurement of the resonator's reflection coefficient is shown in Figure \ref{sup:fig:DR}b, overlaid with a fitting function \cite{probst2015efficient} used to extract the internal ($\kappa_{\rm i}$) and external ($\kappa_{\rm e}$) loss rates that are important in our experiments. An internal quality factor of approximately $Q_{\rm i} = 11,000$ is obtained from the fit.

\subsection{Optics}\label{sup:subsec:optics}
\begin{figure}[h]
\includegraphics{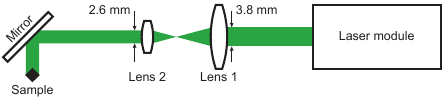}
\caption{\textbf{Optical setup.} Diagram of the optical setup, containing a Lasertack LAB-520-1500 laser module, Thorlabs LA1027-A 35 mm focal length lens (Lens 1), Thorlabs AC254-045-A1-ML 45 mm focal length lens (Lens 2) and Thorlabs DMLP550R dichroic mirror.}
\label{sup:fig:Optics}
\end{figure}

To ensure that we polarize as many of the NV$^-$ spins as possible without illuminating and heating the resonator, we use a beam diameter of 2.6~mm that was carefully chosen to interface with the resonator design and sample dimensions. While the sample has a maximum width of 3.3~mm when viewed from the top, 97.6~\% of the sample volume is covered by a laser beam with a 2.6~mm diameter centered on the sample. The region of the sample that is outside of this beam is used to support the diamond, ensuring that no part of the resonator is in the direct path of the laser beam. The laser diode module outputs a 3.8~mm diameter beam. To achieve the required 2.6~mm diameter beam we use a Keplerian beam reducer and direct the beam using a dichroic mirror fixed to an adjustable mount (Figure~\ref{sup:fig:Optics}).

\subsection{Helium gas thermalization}\label{sup:subsec:He}
\begin{figure}[h]
\includegraphics{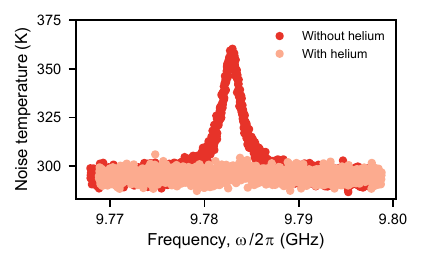}
\caption{\textbf{Microwave noise temperature with spins detuned.} Inferred resonator output noise temperature at maximum laser power (1.61~W), with and without helium gas flowing through the setup.}
\label{sup:fig:Heating}
\end{figure}

Illuminating the diamond sample with the green 520~nm laser causes it to heat, since the only available channels for thermalization are through the surrounding air or the low thermal conductivity dielectric ceramic material (2.56~Wm$^{-1}$K$^{-1}$ at 294~K) and the quartz tube used to support the resonator (see Fig.~1d of the main text). Any stray laser light that hits the resonator will also cause it to heat. To better thermalize the components, we flow room-temperature helium gas through the quartz tube and over the sample and resonator assembly during operation. Helium has a higher thermal conductivity (0.15~Wm$^{-1}$K$^{-1}$) than air (0.026Wm$^{-1}$K$^{-1}$) at room temperature and atmospheric pressure. Without the helium gas we find (using the method outlined in Section~\ref{sup:sec:cooling}) that the temperature of the noise emitted by the resonator (its output noise temperature) increases by more than 60~K (see Fig.~\ref{sup:fig:Heating}) at our maximum applied laser power (1.61~W). A higher noise temperature is detrimental to the device's performance both a maser amplifier and a microwave cooler. Additionally, a higher temperature of the diamond sample lowers the NV$^-$ spin relaxation time $T_1$, limiting the amount of polarization achievable via optical pumping. With the helium gas applied we see the microwave output noise return to room temperature (Fig.~\ref{sup:fig:Heating}) and also observe that the spin $T_1$ time increases to 6.7(1)~ms, up from 2.6(1)~ms without helium (see Section~\ref{sup:subsec:T1T2}).

\subsection{Pulsed electron spin resonance}\label{sup:subsec:ESR_setup}
\begin{figure}[h]
\includegraphics{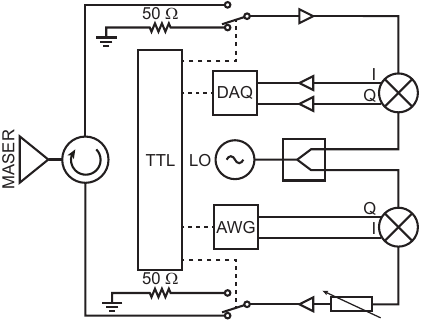}
\caption{\textbf{Pulsed ESR measurement setup.} Schematic of the setup used to perform the pulsed ESR experiments.}
\label{sup:fig:ESRSetup}
\end{figure}

All of the pulsed ESR measurements were obtained using a home-built spectrometer depicted in Fig.~\ref{sup:fig:ESRSetup}. The system has two main arms: one for generating phase-controlled pulses, and one for performing homodyne demodulation of the reflected signals. Pulses are generated from a Local Oscillator (LO) with fixed power using two channels of an Arbitrary Waveform Generator (AWG) with a microwave IQ-mixer. To extend the dynamic range of the system, a programmable attenuator can be used to vary the output power of the system. To ensure no unintended power leaks out of the box, a fast microwave switch is placed before the output of the system and is
actuated throughout the pulse sequences. Similar switches are placed at the input of the box and provide 40 dB of attenuation when open to ensure that the high power pulses used for performing ESR do not reach the demodulator with a power exceeding its damage threshold. Power entering the box is further amplified before being homodyne demodulated. The resulting quadrature signals are digitized by a Data Acquisition System (DAQ). Triggering of the AWG, DAQ, and microwave switches is achieved with TTL logic supplied by Spin-Core PulseBlaster ESR Pro.

\section{Input-output theory for the maser amplifier/cooler system}\label{sup:sec:Theory}
In the following Section we develop an input-output theoretical model to describe the mean field and the second moment of the resonator field (noise) for our system. We follow the formalism of Gardiner and Collett \cite{gardiner1985input} and develop the model for both an inverted spin ensemble (i.e. the amplification regime) and a spin ensemble in its ground state (the cooling regime). This builds on previous work studying resonators coupled with spin ensembles in inverted \cite{julsgaard2012dynamical} and ground state \cite{kurucz2011spectroscopic,diniz2011strongly,fahey2023steady, bienfait2017magnetic} configurations.

\subsection{System Hamiltonian}
The Hamiltonian describing a single resonator mode coupled to an ensemble of NV$^-$ spins, with the resonator mode and spins both coupled to heat baths is:
\begin{equation}
H = H_{\rm r} + H_{\rm rB} + H_{\rm rB,int} + H_{\rm s} + H_{\rm sr,int} + H_{\rm sB} + H_{\rm sB,int}
\end{equation}
where $H_{\rm r}$ is the energy of the resonator mode, $H_{\rm rB}$ is the total energy of the resonator bath, $H_{\rm rB,int}$ describes the interaction between the mode and the bath, $H_{\rm s}$ is the total energy of the spin ensemble, $H_{\rm sr,int}$ is the Jaynes-Cummings interaction between each spin in the ensemble and the resonator mode, $H_{\rm sB}$ is the total energy of the spin bath and $H_{\rm sB,int}$ describes the interaction between the each spin and an independent bath. These Hamiltonians are defined as:
\begin{align}
H_{\rm r} &= \hbar \omega_{\rm r} a^\dagger a\\
H_{\rm rB} &= \hbar \int_{-\infty}^{\infty} \,d\omega\, \omega b^\dagger(\omega)b(\omega)\\
H_{\rm rB,int} &= i\hbar \int_{-\infty}^{\infty} \,d\omega\, \kappa(\omega)\left[ a^\dagger b(\omega) - ab^\dagger(\omega)\right]\\
H_{\rm s} &= \frac{\hbar}{2} \sum_{j=1}^{N} \omega_{j} \sigma_{\rm z}^{(j)}\\
H_{\rm sr,int} &= \hbar \sum_{j=1}^{N} \left( g^*_{j} \sigma_{\rm +}^{(j)} a + g_{j} \sigma_{\rm -}^{(j)} a^\dagger \right)
\end{align}
where $a$ ($a^\dagger$) is the annihilation (creation) operator for the resonator mode (which has a resonance frequency of $\omega_{\rm r}$), $b(\omega)$ ($b^\dagger(\omega)$) is the annihilation (creation) operator of the bath at frequency $\omega$, $\kappa(\omega)$ is the coupling between the mode and the bath, $\sigma_{\rm z}^{(j)}$, $\sigma_{\rm +}^{(j)}$ and $\sigma_{\rm -}^{(j)}$ are, respectively, the Pauli Z, raising and lowering operators describing the $j$-th spin in the ensemble (out of $N$ total spins), which has a resonance frequency of $\omega_j$ and is coupled with a strength $g_j$ to the resonator mode. In the following section we define the spin bath ($H_{\rm sB}$) and its interaction with the spins ($H_{\rm sB,int}$).

\subsection{Quantum Langevin equations for the maser amplifier}\label{sub:sec:qlanmaser}
Under strong optical pumping, the NV$^-$ spins at the maser amplification transition are high polarized in the excited state (see Fig.~2c of main text). We invoke the Holstein-Primakoff approximation \cite{julsgaard2012dynamical,kurucz2011spectroscopic,diniz2011strongly,bienfait2017magnetic}, which is valid for highly polarized spin systems, to model each spin as a harmonic oscillator. For an inverted ensemble the mapping from spin to boson operators is:
\begin{align}
\sigma_{\rm -}^{(j)} &\rightarrow s_j^\dagger\\
\sigma_{\rm +}^{(j)} &\rightarrow s_j\\
\sigma_{\rm z}^{(j)} &\rightarrow 1 - 2s_j^\dagger s_j
\end{align}
where $s_j$ ($s_j^\dagger$) is the annihilation (creation) operator for the j-th spin. This describes an inverted (i.e., upside down) ensemble of harmonic oscillators \cite{gardiner2004quantum}. A spin de-excitation creates an excitation in the inverted oscillator system, which lowers its energy. Under this approximation we can re-write the spin energy and resonator-spin interaction Hamiltonians as:
\begin{align}
H_{\rm s} &= -\hbar \sum_{j=1}^N \omega_j s_j^\dagger s_j\\
H_{\rm sr,int} &= \hbar \sum_{j=1}^{N} \left( g^*_{j} s_j a + g_{j} s_j^\dagger a^\dagger \right)
\end{align}

We make the assumption that each spin interacts with an independent heat bath of inverted harmonic oscillators and define the following Hamiltonians for the total inverted spin bath energy $H_{\rm sB}$ and the interaction of each spin with its independent bath $H_{\rm SB,int}$:
\begin{align}
H_{\rm sB} &= -\hbar \sum_{j=1}^{N} \int_{-\infty}^{\infty} \,d\omega\, \omega d_j^\dagger (\omega)d_j(\omega)\\
H_{\rm SB,int} &= i\hbar \sum_{j=1}^{N} \int_{-\infty}^{\infty} \,d\omega\, \gamma(\omega)\left[ s_j^\dagger d_j(\omega) - s_j d_j^\dagger(\omega)\right]
\end{align}
where $d_j(\omega)$ ($d_j^\dagger (\omega)$) is the annihilation (creation) operator of the bath at frequency $\omega$ for the j-th spin and $\gamma(\omega)$ is the coupling between each spin and its bath.

Next we derive the Heisenberg equations of motion for the system operators:
\begin{alignat}{2}
\dot{b}(\omega) &= \frac{1}{i\hbar} \left[b(\omega),H\right] &&= -i\omega b(\omega) - \kappa(\omega)a \label{sup:eqn:bdot}\\
\dot{a} &= \frac{1}{i\hbar} \left[a,H\right] &&= -i\omega_{\rm r} a + \int_{-\infty}^{\infty}d\omega\, \kappa(\omega)b(\omega) -i\sum_{j=1}^{N} g_j s_j^\dagger \label{sup:eqn:adot}\\
\dot{d_j^\dagger}(\omega) &= \frac{1}{i\hbar} \left[d_j^\dagger,H\right] &&= -i\omega d_j^\dagger(\omega) -\gamma (\omega) s_j^+ \label{sup:eqn:ddot}\\
\dot{s_j^\dagger} &= \frac{1}{i\hbar} \left[s_j^\dagger,H\right] &&= -i\omega_js_j^\dagger +ig^*_ja + \int_{-\infty}^{\infty}d\omega\, \gamma(\omega)d_j^\dagger(\omega) \label{sup:eqn:sdagdot}
\end{alignat}

The solutions for Eqs.~\ref{sup:eqn:bdot} and \ref{sup:eqn:ddot} are given as:
\begin{align}
b(\omega) &= e^{-i\omega(t-t_0)} b_0(\omega) - \kappa(\omega) \int_{t_0}^{t} dt^\prime \, e^{-i\omega(t-t^\prime)}a(t^\prime)\label{sup:eqn:b}\\
d_j^\dagger(\omega) &= e^{-i\omega(t-t_0)} {d_0^{(j)}}^\dagger(\omega) - \gamma(\omega) \int_{t_0}^{t} dt^\prime \, e^{-i\omega(t-t^\prime)}s_j^\dagger(t^\prime)\label{sup:eqn:djdag}
\end{align}

We substitute Eqs.~\ref{sup:eqn:b} and \ref{sup:eqn:djdag} into Eqs.~\ref{sup:eqn:adot} and \ref{sup:eqn:sdagdot} to obtain:
\begin{align}
\dot{a} &= -i \omega_{\rm r} a - i \sum_{j=1}^{N} g_j s_j^\dagger + \int_{-\infty}^{\infty} d\omega\, \kappa(\omega) e^{-i\omega(t-t_0)}b_0(\omega) - \int_{-\infty}^{\infty} d\omega\, \kappa(\omega)^2 \int_{t_0}^{t} dt^\prime \, e^{-i \omega(t-t^\prime)}a(t^\prime)\label{sup:eqn:adot2}\\
\dot{s_j^\dagger} &= -i \omega_{\rm r} s_j^\dagger + ig^*_ja + \int_{-\infty}^{\infty} d\omega\, \gamma(\omega) e^{-i\omega(t-t_0)}{d_0^{(j)}}^\dagger(\omega) - \int_{-\infty}^{\infty} d\omega\, \gamma(\omega)^2 \int_{t_0}^{t} dt^\prime \, e^{-i \omega(t-t^\prime)}s_j^\dagger(t^\prime)\label{sup:eqn:sjdagdot2}
\end{align}

To simplify Eqs.~\ref{sup:eqn:adot2} and \ref{sup:eqn:sjdagdot2} we assume that the coupling constants are independent of frequency, i.e. $\kappa(\omega)=\sqrt{\kappa/(2\pi)}$ and $\gamma(\omega)=\sqrt{\gamma/(2\pi)}$, which is referred to as the first Markov approximation \cite{gardiner1985input}. We also use the following properties:
\begin{align}
\int_{-\infty}^{\infty} d\omega\, e^{-i\omega(t-t^\prime)} &= 2\pi \delta(t-t^\prime)\\
\int_{t_0}^{t} dt^\prime \, X(t^\prime) \delta(t-t^\prime) &= \frac{1}{2}X(t)
\end{align}

Furthermore, we define the fields:
\begin{align}
b_{\rm in}(t) &= \frac{1}{\sqrt{2\pi}} \int_{-\infty}^{\infty}d\omega\, e^{-i\omega(t-t_0)}b_0(\omega)\\
{d_{\rm in}^{(j)}}^\dagger(t) &= \frac{1}{\sqrt{2\pi}} \int_{-\infty}^{\infty}d\omega\, e^{-i\omega(t-t_0)}{d_0^{(j)}}^\dagger(\omega)
\end{align}

Combining these approximations, properties and definitions, we can rewrite Eqs.~\ref{sup:eqn:adot2} and \ref{sup:eqn:sjdagdot2} as:
\begin{align}
\dot{a} &= -i\omega_{\rm r} a - i\sum_{j=1}^{N}g_j s_j^\dagger + \sqrt{\kappa}\,b_{\rm in}(t) - \frac{\kappa}{2}a\label{sup:eqn:adot3}\\
\dot{s_j^\dagger} &= -i\omega_j s_j^\dagger + i g^*_ja +\sqrt{\gamma}\,{d_{\rm in}^{(j)}}^\dagger(t) - \frac{\gamma}{2}s_j^\dagger\label{sup:eqn:sjdagdot3}
\end{align}

In practice, the resonator mode is coupled to two independent baths, one representing the internal channels of loss (coupled at a rate $\kappa_{\rm i}$) and the other representing modes in the external loop coupler (coupled at a rate $\kappa_{\rm e}$). We thus rewrite Eq.~\ref{sup:eqn:adot3} as:
\begin{equation}
\dot{a} = -i\omega_{\rm r} a - i\sum_{j=1}^{N}g_j s_j^\dagger + \sqrt{\kappa_{\rm e}}\,a_{\rm in}(t) - \frac{\kappa_{\rm e}}{2}a + \sqrt{\kappa_{\rm i}}\,b_{\rm in}(t) - \frac{\kappa_{\rm i}}{2}a\label{sup:eqn:adot4}
\end{equation}
where $a_{\rm in}(t)$ is the input field from the loop coupler (i.e. the field travelling to the resonator via the coupler) and $b_{\rm in}(t)$ is the input field from the resonator heat bath.

We define the Fourier-transformed operator:
\begin{equation}
a[\omega] = \frac{1}{\sqrt{2\pi}} \int_{-\infty}^{\infty}\,dt\,e^{i\omega t}a(t)\label{sup:eqn:ft}
\end{equation}
and write Eqs.~\ref{sup:eqn:sjdagdot3} and \ref{sup:eqn:adot4} in the Fourier domain:
\begin{align}
-i\omega a[\omega] &= -i\omega_{\rm r} a[\omega] -i \sum_{j=1}^{N} g_j s_j^\dagger[\omega] + \sqrt{\kappa_{\rm e}}\,a_{\rm in}[\omega] + \sqrt{\kappa_{\rm i}}\,b_{\rm in}[\omega] - \bar{\kappa}a[\omega]\label{sup:eqn:aw}\\
-i\omega s_j^\dagger[\omega] &= -i\omega_j s_j^\dagger[\omega] + i g^*_ja[\omega] + \sqrt{\gamma}\,{d_{\rm in}^{(j)}}^\dagger[\omega] - \frac{\gamma}{2}s_j^\dagger[\omega]\label{sup:eqn:sjdagw}
\end{align}
where $\bar{\kappa}=(\kappa_{\rm e}+\kappa_{\rm i})/2$.

Next we solve Eq.~\ref{sup:eqn:sjdagw} for $ s_j^\dagger[\omega]$ and substitute the result into Eq.~\ref{sup:eqn:aw}. We also apply the input-output relation $a_{\rm out}[\omega] + a_{\rm in}[\omega] = \sqrt{\kappa_e}\,a[\omega]$ which relates the intra-resonator field $a[\omega]$ to the input $a_{\rm in}[\omega]$ and output $a_{\rm out}[\omega]$ fields inside the coupler, giving:
\begin{equation}
a_{\rm out}[\omega] = \left( \frac{i\kappa_{\rm e}}{\omega-\omega_{\rm r} +i\bar{\kappa} +K(\omega)} -1 \right)a_{\rm in}[\omega] + \frac{i \sqrt{\kappa_{\rm e}}}{\omega-\omega_{\rm r}+i\bar{\kappa}+K(\omega)} \sum_{j=1}^{N} \frac{\sqrt{\gamma}g_j{d_{\rm in}^{(j)}}^\dagger[\omega]}{\omega-\omega_j+i\frac{\gamma}{2}} + \frac{i\sqrt{\kappa_{\rm e}\kappa_{\rm i}}}{\omega-\omega_{\rm r}+i\bar{\kappa}+K(\omega)}b_{\rm in}[\omega]\label{sup:eqn:aoutw}
\end{equation}

where,
\begin{equation}
K(\omega) = \sum_{j=1}^{N} \frac{|g_j|^2}{\omega-\omega_j+i\frac{\gamma}{2}}\label{sup:eqn:kw}
\end{equation}

Equation~\ref{sup:eqn:aoutw} expresses the output field (the field inside the coupler travelling away from the resonator) in terms of the input field and the resonator and spin baths. From this expression we can calculate the mean fields and second moments to find expressions for the maser amplifier gain and noise.

From the definition of the Fourier-transformed operator (Eq.~\ref{sup:eqn:ft}), we see:
\begin{equation}
{a[\omega]}^\dagger = \frac{1}{\sqrt{2\pi}} \int_{-\infty}^{\infty}\,dt\,e^{-i\omega t}a^\dagger(t) = a^\dagger[-\omega]
\end{equation}
i.e. taking the adjoint of the operator $a[\omega]$ in the Fourier domain is equivalent to taking the Fourier transform of $a^\dagger(t)$ and evaluating the result at $-\omega$. Equivalently, we can show that ${d_{\rm in}^{(j)}}^\dagger[\omega] = {d_{\rm in}^{(j)}[-\omega]}^\dagger$ and thus ${{d_{\rm in}^{(j)}}^\dagger[\omega]}^\dagger = d_{\rm in}^{(j)}[-\omega]$.

\subsubsection{Reflection parameter}
In our experiments we probe the system with a coherent microwave tone $\langle a_{\rm in}[\omega]\rangle$ and measure the reflected signal $\langle a_{\rm out}[\omega]\rangle$, which allows us to determine the reflection coefficient $r$:
\begin{equation} \label{sup:eqn:r}
r = \frac{\langle a_{\rm out}[\omega]\rangle}{\langle a_{\rm in}[\omega]\rangle} = \frac{i\kappa_{\rm e}}{\omega - \omega_{\rm r} + i\bar{\kappa} + K\left(\omega\right)} -1
\end{equation}

Here we make the assumption that all fields (aside from $a_{\rm in}[\omega]$) are in thermal states, such that $\langle {d_{\rm in}^{(j)}[-\omega]}^\dagger\rangle = \langle b_{\rm in}[\omega]\rangle = 0$.

\subsubsection{Output noise}\label{sup:sec:nout}
To determine the noise in the output field $a_{\rm out}[\omega]$, we first define the output quadrature operators:
\begin{align} 
I_{\rm out} &= \frac{1}{2}\left({a_{\rm out}[\omega]}^\dagger + a_{\rm out}[\omega]\right)\label{sup:eqn:iout}\\
Q_{\rm out} &= \frac{i}{2}\left({a_{\rm out}[\omega]}^\dagger - a_{\rm out}[\omega]\right)\label{sup:eqn:qout}
\end{align}

We assume that all fields are in thermal states $\langle a_{\rm in}[\omega]\rangle = \langle {d_{\rm in}^{(j)}[-\omega]}^\dagger\rangle = \langle b_{\rm in}[\omega]\rangle = 0$. The second moment of the output field can be calculated by summing the moments of the quadrature operators:
\begin{align}
n_{\rm out} = &\langle I_{\rm out}^2\rangle + \langle Q_{\rm out}^2\rangle\\
= &\frac{1}{4} \left< ({a_{\rm out}[\omega]}^\dagger + a_{\rm out}[\omega])^2\right> - \frac{1}{4}\left<({a_{\rm out}[\omega]}^\dagger - a_{\rm out}[\omega])^2 \right>\\
= &\frac{1}{2}\left<{a_{\rm out}[\omega]}^\dagger a_{\rm out}[\omega] + a_{\rm out}[\omega]{a_{\rm out}[\omega]}^\dagger\right>\label{sup:eqn:nout}
\end{align}

Using Eq.~\ref{sup:eqn:aoutw}, we therefore find:
\begin{equation}
n_{\rm out} = R_{\rm in}\left(n_{\rm in} + \frac{1}{2}\right) +
R_{\rm s}\left(n_{\rm s} + \frac{1}{2}\right) +
R_{\rm i}\left(n_{\rm i} + \frac{1}{2}\right)\label{sup:eqn:nout2}
\end{equation}
where $n_{\rm in} = \langle {a_{\rm in}[\omega]}^\dagger a_{\rm in}[\omega]\rangle$, $n_{\rm s} = \langle{d_{\rm in}^{(j)}[-\omega]}^\dagger{d_{\rm in}^{(j)}[-\omega]}\rangle$, $n_{\rm i} = \langle {b_{\rm in}[\omega]}^\dagger b_{\rm in}[\omega]\rangle$ are the thermal occupations of the various fields (given by Bose-Einstein statistics), and:
\begin{align}
R_{\rm in} &=  \left|\frac{i\kappa_{\rm e}}{\omega - \omega_{\rm r} + i\bar{\kappa} + K\left(\omega\right)} -1\right| = |r|^2\\
R_{\rm s} &= \frac{\kappa_{\rm e}}{\left|\omega - \omega_{\rm r} + i\bar{\kappa} + K\left(\omega\right)\right|^2}C(\omega)\\
R_{\rm i} &= \frac{\kappa_{\rm e}\kappa_{\rm i}}{\left|\omega - \omega_{\rm r} + i\bar{\kappa} + K\left(\omega\right)\right|^2}
\end{align}
We further define the function:
\begin{equation}
C(\omega) = 2\sum_{j=1}^{N} \frac{|g_j|^2\frac{\gamma}{2}}{\left(\omega-\omega_j\right)^2+\left(\frac{\gamma}{2}\right)^2}\label{sup:eqn:cw}
\end{equation}

Later we will further explore the meaning of the functions $K(\omega)$ and $C(\omega)$.

To obtain Eqs.~\ref{sup:eqn:nout2}-\ref{sup:eqn:cw}, we have made the assumption that the different fields are uncorrelated, i.e. $\langle {a_{\rm in}[\omega]}^\dagger b_{\rm in}[\omega]\rangle = 0$, $\langle {a_{\rm in}[\omega]}^\dagger d_{\rm in}^{(j)}[-\omega]\rangle = 0$, $\langle {b_{\rm in}[\omega]}^\dagger d_{\rm in}^{(j)}[-\omega]\rangle = 0$, $\langle {d_{\rm in}^{(j)}[-\omega]}^\dagger d_{\rm in}^{(j+1)}[-\omega]\rangle = 0$, $\langle a_{\rm in}[\omega] b_{\rm in}[\omega]\rangle = 0$, etc.

\subsubsection{Input-referred noise}
We are typically interested in the equivalent noise that is added to a signal before it is amplified. This is found by dividing the output noise by the gain of the amplifier ($R_{\rm in}$), giving the input-referred noise:
\begin{equation}
\frac{n_{\rm out}}{R_{\rm in}} = \left(n_{\rm in} + \frac{1}{2}\right) + \frac{R_{\rm s}}{R_{\rm in}}\left(n_{\rm s} + \frac{1}{2}\right) + \frac{R_{\rm i}}{R_{\rm in}}\left(n_{\rm i} + \frac{1}{2}\right)\label{sup:eqn:noutonrin}
\end{equation}

where:
\begin{align}
\frac{R_{\rm s}}{R_{\rm in}} &= \frac{\kappa_{\rm e}}{\left|\omega_{\rm r} - \omega + i\frac{\kappa_{\rm e} - \kappa_{\rm i}}{2} - K\left(\omega\right)\right|^2}C(\omega)\\
\frac{R_{\rm i}}{R_{\rm in}} &= \frac{\kappa_{\rm e}\kappa_{\rm i}}{\left|\omega_{\rm r} - \omega + i\frac{\kappa_{\rm e} - \kappa_{\rm i}}{2} - K\left(\omega\right)\right|^2}
\end{align}

The first term in Eq.~\ref{sup:eqn:noutonrin} represents noise already on the signal, which is not intrinsic to the maser amplifier. We therefore define the input-referred noise of the maser amplifier as:
\begin{equation}
n_{\rm m} = \frac{n_{\rm out}}{R_{\rm in}} - \left(n_{\rm in} + \frac{1}{2}\right) = \frac{R_{\rm s}}{R_{\rm in}}\left(n_{\rm s} + \frac{1}{2}\right) + \frac{R_{\rm i}}{R_{\rm in}}\left(n_{\rm i} + \frac{1}{2}\right)
\end{equation}

\subsection{Quantum Langevin equations for the microwave cooler}
Under strong optical pumping, the NV$^-$ spins at the microwave cooling transition are highly polarized in their ground state. In this case the Holstein-Primakoff approximation \cite{kurucz2011spectroscopic,diniz2011strongly}, provides a different mapping of the spin operators to boson operators:
\begin{align}
\sigma_{\rm -}^{(j)} &\rightarrow s_j\\
\sigma_{\rm +}^{(j)} &\rightarrow s_j^\dagger\\
\sigma_{\rm z}^{(j)} &\rightarrow 2s_j^\dagger s_j - 1
\end{align}

This describes a system of regular (i.e., non-inverted) harmonic oscillators and results in the following spin Hamiltonians:
\begin{align}
H_{\rm s} &= \hbar \sum_{j=1}^N \omega_j s_j^\dagger s_j\\
H_{\rm sr,int} &= \hbar \sum_{j=1}^{N} \left( g^*_{j} s_j^\dagger a + g_{j} s_j a^\dagger \right)
\end{align}

We again make the assumption that each spin interacts with an independent heat bath, however, this time consisting of regular harmonic oscillators. We define the following Hamiltonians for the total spin bath energy $H_{\rm sB}$ and the interaction of each spin with its independent bath $H_{\rm SB,int}$:
\begin{align}
H_{\rm sB} &= \hbar \sum_{j=1}^{N} \int_{-\infty}^{\infty} \,d\omega\, \omega d_j^\dagger (\omega)d_j(\omega)\\
H_{\rm SB,int} &= i\hbar \sum_{j=1}^{N} \int_{-\infty}^{\infty} \,d\omega\, \gamma(\omega)\left[ s_j^\dagger d_j(\omega) - s_j d_j^\dagger(\omega)\right]
\end{align}

All other Hamiltonians are identical to those presented in Section~\ref{sub:sec:qlanmaser}. We follow the same procedure as in Section~\ref{sub:sec:qlanmaser} to write down the quantum Langevin equations for the microwave cooling regime and find the following expression for the output field:
\begin{equation}
a_{\rm out}[\omega] = \left( \frac{i\kappa_{\rm e}}{\omega-\omega_{\rm r} +i\bar{\kappa} - K(\omega)} -1 \right)a_{\rm in}[\omega] + \frac{i \sqrt{\kappa_{\rm e}}}{\omega-\omega_{\rm r}+i\bar{\kappa} - K(\omega)} \sum_{j=1}^{N} \frac{\sqrt{\gamma}g_j d_{\rm in}^{(j)}[\omega]}{\omega-\omega_j+i\frac{\gamma}{2}} + \frac{i\sqrt{\kappa_{\rm e}\kappa_{\rm i}}}{\omega-\omega_{\rm r}+i\bar{\kappa} - K(\omega)}b_{\rm in}[\omega]\label{sup:eqn:aoutwc}
\end{equation}

We note two main differences between Eqs.~\ref{sup:eqn:aoutw} and \ref{sup:eqn:aoutwc}. The first is that the spin bath field operator ${d_{\rm in}^{(j)}}^\dagger[\omega]$ has become $d_{\rm in}^{(j)}[\omega]$ in Eq.~\ref{sup:eqn:aoutwc}. The second difference is a sign change in front of the function $K(\omega)$ in the denominator of each of the terms. These changes ensure that the commutator of the output field operator $[a_{\rm out},a_{\rm out}^\dagger]=1$ is obeyed for both amplification and cooling.

\subsubsection{Reflection parameter}
We find the reflection parameter for the system operated in the cooling regime to be:
\begin{equation} \label{sup:eqn:r}
r = \frac{\langle a_{\rm out}[\omega]\rangle}{\langle a_{\rm in}[\omega]\rangle} = \frac{i\kappa_{\rm e}}{\omega - \omega_{\rm r} + i\bar{\kappa} - K\left(\omega\right)} -1
\end{equation}

\subsubsection{Output noise}
The output noise in the cooling regime is calculated using the same steps as presented in Section~\ref{sup:sec:nout}:
\begin{equation}
n_{\rm out} = R_{\rm in}\left(n_{\rm in} + \frac{1}{2}\right) +
R_{\rm s}\left(n_{\rm s} + \frac{1}{2}\right) +
R_{\rm i}\left(n_{\rm i} + \frac{1}{2}\right)\label{sup:eqn:nout3}
\end{equation}
where $n_{\rm in} = \langle {a_{\rm in}[\omega]}^\dagger a_{\rm in}[\omega]\rangle$, $n_{\rm s} = \langle{d_{\rm in}^{(j)}[\omega]}^\dagger{d_{\rm in}^{(j)}[\omega]}\rangle$, $n_{\rm i} = \langle {b_{\rm in}[\omega]}^\dagger b_{\rm in}[\omega]\rangle$ are the thermal occupations of the various fields (given by Bose-Einstein statistics), and:
\begin{align}
R_{\rm in} &=  \left|\frac{i\kappa_{\rm e}}{\omega - \omega_{\rm r} + i\bar{\kappa} - K\left(\omega\right)} -1\right| = |r|^2\\
R_{\rm s} &= \frac{\kappa_{\rm e}}{\left|\omega - \omega_{\rm r} + i\bar{\kappa} - K\left(\omega\right)\right|^2}C(\omega)\\
R_{\rm i} &= \frac{\kappa_{\rm e}\kappa_{\rm i}}{\left|\omega - \omega_{\rm r} + i\bar{\kappa} - K\left(\omega\right)\right|^2}
\end{align}

\subsection{Unified expressions for amplification and cooling}\label{sup:subsec:unified}
In the previous sections we showed that the reflection and output noise of the system, operated either in the amplification or cooling regime, are identical apart from a sign change of the $K(\omega)$ function. We can therefore express the reflection parameter in the succinct form:
\begin{equation}
r^\pm = \frac{i\kappa_{\rm e}}{\omega - \omega_{\rm r} + i\bar{\kappa} \pm K\left(\omega\right)} -1\label{sup:eqn:rpm}
\end{equation}
where $+$ represents amplification (for an inverted spin ensemble) and $-$ signifies cooling (for a spin ensemble polarized in its ground state).

We can further express the output noise as:
\begin{equation}
n_{\rm out}^\pm = R_{\rm in}^\pm\left(n_{\rm in} + \frac{1}{2}\right) +
R_{\rm s}^\pm\left(n_{\rm s} + \frac{1}{2}\right) +
R_{\rm i}^\pm\left(n_{\rm i} + \frac{1}{2}\right)
\end{equation}
where:
\begin{align}
R_{\rm in}^\pm &=  \left|\frac{i\kappa_{\rm e}}{\omega - \omega_{\rm r} + i\bar{\kappa} \pm K\left(\omega\right)} -1\right| = |r^\pm|^2\label{sup:eqn:rinpm}\\
R_{\rm s}^\pm &= \frac{\kappa_{\rm e}}{\left|\omega - \omega_{\rm r} + i\bar{\kappa} \pm K\left(\omega\right)\right|^2}C(\omega)\\
R_{\rm i}^\pm &= \frac{\kappa_{\rm e}\kappa_{\rm i}}{\left|\omega - \omega_{\rm r} + i\bar{\kappa} \pm K\left(\omega\right)\right|^2}\label{sup:eqn:ripm}
\end{align}

These are the equations presented in the main text. 

\subsection{Bath thermal occupations}\label{sup:subsec:baths}
Each bath is assumed to be in equilibrium at an effective temperature $T_x$ (where $x = \{\text{in, s, i}\})$, with the number of thermal photons $n_x$ populating each bath given by the Bose-Einstein distribution:
\begin{equation}
n_x = \frac{1}{\exp{[\hbar\omega/(k_{\rm B}T_x)]} - 1}\label{sup:eqn:nxbe}
\end{equation}

For the input field, the effective temperature is given by the physical temperature of the input cable ($T_x=T_{\rm in}=294$~K). For the resonator loss we take the physical temperature of the resonator ($T_x=T_{\rm i}=294$~K). The spin bath temperature, on the other hand, depends on the degree of spin polarization induced by optical pumping:
\begin{equation}
\rho = \frac{N_{\rm g}-N_{\rm e}}{N_{\rm g}+N_{\rm e}}\label{sup:eqn:rhoopt}
\end{equation}
where $N_{\rm g}$ and $N_{\rm e}$ correspond to the number of spins in the ground and excited states of the two-level subsystem of the NV$^-$ triplet ground state, respectively. Population inversion in the amplification regime means $N_{\rm e} > N_{\rm g}$ and consequently a negative $\rho$, whilst for cooling $N_{\rm g} > N_{\rm e}$ and $\rho$ is positive. The spin temperature is defined by the relation:
\begin{equation}
\rho = \tanh \left(\frac{\hbar\omega_{\rm s}}{2k_{\rm B}T_{\rm s}}\right)\label{sup:eqn:rhotanh}
\end{equation}

We therefore obtain a negative temperature $T_x=-|T_{\rm s}|$ for amplification and a positive temperature $T_x=|T_{\rm s}|$ for cooling. 

The thermal occupation of the input field ($n_{\rm in} = \langle {a_{\rm in}[\omega]}^\dagger a_{\rm in}[\omega]\rangle$), resonator bath ($n_{\rm i} = \langle {a_{\rm i}[\omega]}^\dagger a_{\rm i}[\omega]\rangle$), and spin bath in the cooling regime ($n_{\rm s} = \langle {a_{\rm s}[\omega]}^\dagger a_{\rm s}[\omega]\rangle$) are all determined by positive temperatures and positive frequencies, thus evaluating Eq.~\ref{sup:eqn:nxbe} is straight-forward. The spin bath in the inverted (amplification) regime is characterised by a negative spin temperature $-|T_{\rm s}|$ and a negative frequency (since $n_{\rm s} = \langle{d_{\rm in}^{(j)}[-\omega]}^\dagger{d_{\rm in}^{(j)}[-\omega]}\rangle$), in this case we find:
\begin{equation}
n_{\rm s} = \frac{1}{\exp{[\hbar(-\omega)/(k_{\rm B}(-|T_{\rm s}|))]} - 1} = \frac{1}{\exp{[\hbar\omega/(k_{\rm B}|T_{\rm s}|)]} - 1}
\end{equation}
which is an identical form to Eq.~\ref{sup:eqn:nxbe}. Thus, the thermal occupations of the inverted and non-inverted spin baths can be calculated in the same way.

\subsection{Spin functions $K(\omega)$ and $C(\omega)$}\label{sup:sec:kc}
Eqs.~\ref{sup:eqn:rpm}-\ref{sup:eqn:ripm} for the most part involve straight-forward system parameters that are easy to determine, aside from the functions $K(\omega)$ and $C(\omega)$  defined in Eqs.~\ref{sup:eqn:kw} and \ref{sup:eqn:cw}, respectively. These functions relate to properties of the spin frequency and coupling strength distributions and for certain profiles have simple analytical forms.

Taking the spin distribution to be a continuous function, we can rewrite Eq.~\ref{sup:eqn:kw} as:
\begin{align}
K(\omega) &= g_{\rm ens}^2\int_{-\infty}^{\infty}d\omega^\prime\, \frac{f(\omega^\prime)}{\omega - \omega^\prime + i\frac{\gamma}{2}} \label{sup:eqn:kw2}
\end{align}
where $f(\omega)$ is the normalized spin resonance frequency distribution \cite{diniz2011strongly}:
\begin{equation}
f(\omega) = \frac{1}{g_{\rm ens}^2}\sum_{j=1}^{N} |g_j|^2\delta(\omega-\omega_j)
\end{equation}
with $\int_{-\infty}^{\infty}d\omega\,f(\omega) = 1$, and where:
\begin{equation}
g_{\rm ens} = \sqrt{\sum_{j=1}^{N} |g_j|^2}
\end{equation}
is the collective coupling strength between the spin ensemble and resonator. Note, for a homogeneous mode profile, where every spin has the same coupling strength $g_0$, this becomes $g_{\rm ens} = \sqrt{N}|g_0|$.

We can similarly rewrite Eq.~\ref{sup:eqn:cw}, assuming a continuous spin distribution:
\begin{align}
C(\omega) &= 2\pi g_{\rm ens}^2\int_{-\infty}^{\infty}d\omega^\prime\, f(\omega^\prime)\frac{1}{\pi}\frac{\frac{\gamma}{2}}{\left(\omega - \omega^\prime\right)^2 + \left(\frac{\gamma}{2}\right)^2} \label{sup:eqn:cw2}
\end{align}
which represents a convolution of the spin resonance frequency distribution and a Lorentzian function of characteristic width $\gamma$. Both functions $K(\omega)$ and $C(\omega)$ can be solved numerically for arbitrary spin resonance frequency distributions. However, for some of the typical profiles encountered, analytical expressions also exist, as presented below.

Finally, we define the characteristic width of the spin distribution $\Gamma_{\rm eff}$ as:
\begin{equation}
\Gamma_{\rm eff}^{-1} = \frac{1}{2}\int_{-\infty}^{\infty}d\omega^\prime\, \frac{f(\omega^\prime)}{i(\omega^\prime - \omega_{\rm s}) + \frac{\gamma}{2}} = \frac{i}{2 g_{\rm ens}^2}K(\omega_{\rm s}) \label{sup:eqn:gameff} 
\end{equation}

\subsubsection{Lorentzian profile}
For a Lorentzian spin frequency distribution, we have:
\begin{equation}
f(\omega) = \frac{1}{\pi}\frac{\frac{\Gamma}{2}}{(\omega - \omega_{\rm s})^2 + \left(\frac{\Gamma}{2}\right)^2}\label{sup:eqn:fwl}
\end{equation}
where $\omega_{\rm s}$ is the center spin frequency of the distribution and $\Gamma$ characterizes the inhomogeneous broadening of the ensemble. Substituting Eq.~\ref{sup:eqn:fwl} into \ref{sup:eqn:kw2}, one can show \cite{kurucz2011spectroscopic,diniz2011strongly}:
\begin{equation}
K(\omega) = \frac{g_{\rm ens}^2}{(\omega - \omega_{\rm s}) + i\frac{\Gamma+\gamma}{2}} \label{sup:eqn:kw2l}
\end{equation}

By substituting Eq.~\ref{sup:eqn:fwl} into \ref{sup:eqn:cw2}, we also find:
\begin{equation}
C(\omega) = 2 g_{\rm ens}^2\frac{\frac{\Gamma+\gamma}{2}}{(\omega - \omega_{\rm s})^2 + \left(\frac{\Gamma+\gamma}{2}\right)^2} \label{sup:eqn:cw2l}
\end{equation}

At resonance with the spins ($\omega = \omega_{\rm s}$), these functions simplify to:
\begin{align}
K(\omega_{\rm s}) &= -i\frac{2 g_{\rm ens}^2}{\Gamma_{\rm eff}} = -i\frac{\kappa_{\rm s}}{2}\\
C(\omega_{\rm s}) &= \frac{4 g_{\rm ens}^2}{\Gamma_{\rm eff}} = \kappa_{\rm s}
\end{align}
with $\Gamma_{\rm eff} = \Gamma+\gamma$ the effective characteristic width of the spin distribution, and where we have defined a new parameter $\kappa_s = 4g_\text{ens}^2/\Gamma_{\rm eff}$, which describes the emission (absorption) rate of photons to (from) the resonator by the spin ensemble.

For the cooling experiments we observe the hyperfine-split electron spin resonance spectrum of the $m_{\rm s} = 0 \leftrightarrow m_{\rm s} = +1$ spin transition (see Fig.~\ref{sup:fig:Linewidths}a below). The spectrum is well described by a triple Lorentzian profile, which can be modeled by the following spin frequency distribution \cite{grezes2016towards}:
\begin{equation}
f(\omega) = \frac{1}{3}\sum_{j=-1,0,+1}\frac{1}{\pi}\frac{\frac{\Gamma}{2}}{(\omega - (\omega_{\rm s}+j\Delta_{\rm hf}))^2 + \left(\frac{\Gamma}{2}\right)^2}
\end{equation}
where $\Delta_{\rm hf}/2\pi = 2.17$~MHz is the hyperfine splitting.

\subsubsection{Gaussian profile}
If the spin resonance frequencies follow a Gaussian distribution, as observed in our experiments for the maser amplification $m_{\rm s} = 0 \leftrightarrow m_{\rm s} = -1$ spin transition (see Fig.~\ref{sup:fig:Linewidths}b), we have:
\begin{equation}
f(\omega) = \frac{1}{\sqrt{2\pi}}\frac{1}{\sigma}\exp{\left[-\frac{(\omega - \omega_{\rm s})^2}{2\sigma^2}\right]}\label{sup:eqn:fwg}
\end{equation}
where $\sigma$ is the standard deviation of the Gaussian distribution and characterizes the inhomogeneous spin resonance frequency broadening.

Substituting Eq.~\ref{sup:eqn:fwg} into \ref{sup:eqn:kw2}, we find: \cite{kurucz2011spectroscopic,diniz2011strongly}:
\begin{equation}
K(\omega) = \sqrt{\frac{\pi}{2}}\frac{g_{\rm ens}^2}{\sigma}\exp{\left(-\xi^2\right)}\left[\text{erfi}\left(\xi\right)-i\right]\label{sup:eqn:kw2g}
\end{equation}
where $\xi = (\omega - \omega_{\rm s} + i\gamma/2)/(\sqrt{2}\sigma)$ and $\text{erfi}\left(\xi\right)$ is the imaginary error function.

Putting Eq.~\ref{sup:eqn:fwg} into \ref{sup:eqn:cw2} and performing the change of variable $\omega^\prime \rightarrow \omega^\prime + \omega_{\rm s}$, we obtain:
\begin{align}
C(\omega) &= 2 g_{\rm ens}^2\int_{-\infty}^{\infty}d\omega^\prime\, \frac{1}{\sqrt{2\pi}}\frac{1}{\sigma}\exp{\left[-\frac{{\omega^\prime}^2}{2\sigma^2}\right]}\frac{1}{\pi}\frac{\frac{\gamma}{2}}{\left[\left(\omega - \omega_{\rm s}\right) - \omega^\prime\right]^2 + \left(\frac{\gamma}{2}\right)^2}\\
&= 2 g_{\rm ens}^2 \int_{-\infty}^{\infty}d\omega^\prime\, G\left(\omega^\prime;\sigma\right) L\left(\left(\omega - \omega_{\rm s}\right)-\omega^\prime;\gamma\right)\\
&= 2 g_{\rm ens}^2 V\left(\omega - \omega_{\rm s};\sigma;\gamma\right) \label{sup:eqn:cw2g}
\end{align}
where $G\left(\omega;\sigma\right)$ is a normalized Gaussian function with a standard deviation of $\sigma$, $L\left(\omega;\gamma\right)$ is a normalized Lorentzian function with a full width at half maximum (FWHM) of $\gamma$ and $V\left(\omega;\sigma;\gamma\right)$ is a Voigt profile.

We can simplify $K(\omega)$ and $C(\omega)$ by making the assumption that $2\sqrt{2\ln{2}}\sigma \gg \gamma$, i.e. the inhomogeneous broadening of the spin ensemble (taken here as the FWHM $2\sqrt{2\ln{2}}\sigma$) is greater than the single-spin homogeneous broadening ($\gamma$). We estimate the homogeneous broadening as the spin Hahn-echo decoherence rate \cite{grezes2016towards} (see Fig.~\ref{sup:fig:T1_T2} for a Hahn-echo measurement) $\gamma = 2T_2^{-1} = 2\pi\times 0.18$~MHz, which indeed is much less than the measured inhomogeneous broadening at the amplification transition (see Fig.~\ref{sup:fig:Linewidths}a for ESR spectrum) $2\sqrt{2\ln{2}}\sigma = 2\pi\times 3.3$~MHz. Recognizing that for small $\gamma$, the Lorentzian function approaches the Dirac delta function:
\begin{equation}
\delta\left(\omega - \omega^\prime\right) = \lim_{\frac{\gamma}{2}\rightarrow0}{\frac{1}{\pi}\frac{\frac{\gamma}{2}}{\left(\omega - \omega^\prime\right)^2 + \left(\frac{\gamma}{2}\right)^2}}
\end{equation}

From Eq.~\ref{sup:eqn:cw2}, we then find:
\begin{equation}
C(\omega) = 2\pi g_{\rm ens}^2f(\omega)
\end{equation}

At resonance with the spins ($\omega = \omega_{\rm s}$), $K(\omega)$ and $C(\omega)$ thus again simplify to: 
\begin{align}
K(\omega_{\rm s}) &= -i \frac{2g_{\rm ens}^2}{\Gamma_{\rm eff}} = -i\frac{\kappa_{\rm s}}{2}\\
C(\omega_{\rm s}) &= 4\frac{g_{\rm ens}^2}{\Gamma_{\rm eff}} = \kappa_{\rm s}
\end{align}
where the characteristic line width for the Gaussian spin resonance frequency distribution is $\Gamma_{\rm eff} = 2\sqrt{2/\pi}\sigma$.

\section{Maser amplifier noise temperature measurements}\label{sup:sec:noise}
\begin{figure}[h]
\includegraphics{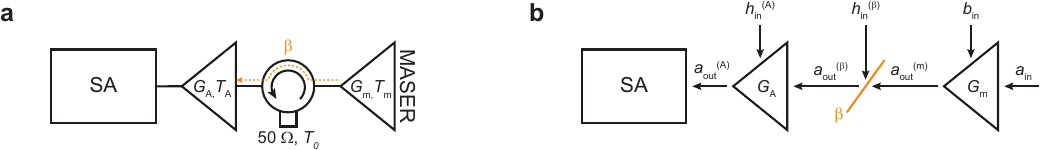}
\caption{\textbf{Amplifier noise temperature measurement setup.} \textbf{a}, Schematic of the setup used to measure the noise temperature of the maser amplifier. \textbf{b}, A simplified schematic of the setup depicting the fields entering and leaving each component.}
\label{sup:fig:NoiseSetup}
\end{figure}

We determine the noise temperature of the maser amplifier $T_{\rm m}$ by comparing the noise power measured with a spectrum analyzer (SA) at the output of the detection path, both with and without the maser amplifier enabled. When the maser amplifier is enabled, it amplifies the room-temperature noise present in the input field, along with the noise introduced by the maser amplifier itself. Any insertion loss between the maser and next amplifier in the chain will also introduce noise, whilst the second amplifier adds further noise to the field. By accurately measuring the maser gain, insertion loss and noise temperature of the second amplifier, we can extract the noise temperature of the maser amplifier.

Figure~\ref{sup:fig:NoiseSetup}a shows the key components of our noise temperature measurement setup. The insertion loss between the maser and second-stage amplifier (resulting from the circulator/isolator and cables) is characterized by a transmissivity $\beta$. We measure the total insertion loss to be $-10\log{\beta}=0.52(10)$~dB, or $\beta = 0.89(4)$. Figure~\ref{sup:fig:NoiseSetup}b portrays a simplified schematic of the setup, depicting the relevant input, output and bath fields for the various elements. The field leaving the maser amplifier $a_{\rm out}^{\rm (m)}$ is related to the input field $a_{\rm in}$ and bath fields ${d_{\rm in}^{(j)}}^\dagger$ and $b_{\rm in}$ via Eq.~\ref{sup:eqn:aoutw}:
\begin{equation}
a_{\rm out}^{\rm (m)} = \left( \frac{i\kappa_{\rm e}}{\omega-\omega_{\rm r} +i\bar{\kappa} +K(\omega)} -1 \right)a_{\rm in} + \frac{i \sqrt{\kappa_{\rm e}}}{\omega-\omega_{\rm r}+i\bar{\kappa}+K(\omega)} \sum_{j=1}^{N} \frac{\sqrt{\gamma}g_j{d_{\rm in}^{(j)}}^\dagger}{\omega-\omega_j+i\frac{\gamma}{2}} + \frac{i\sqrt{\kappa_{\rm e}\kappa_{\rm i}}}{\omega-\omega_{\rm r}+i\bar{\kappa}+K(\omega)}b_{\rm in}\label{sup:eqn:aoutm}
\end{equation}
where we have dropped the frequency dependence of the fields for brevity. 

Loss is modeled as a beam-splitter, which attenuates the maser output field by an amount $\sqrt{\beta}$ and mixes in noise from a thermal bath $h_{\rm in}^{\rm (\beta)}$ at the temperature of the loss. The field at the output of the loss $a_{\rm out}^{\rm (\beta)}$ is therefore \cite{walls2008quantum}:
\begin{equation}
a_{\rm out}^{\rm (\beta)} = \sqrt{\beta}a_{\rm out}^{\rm (m)} + \sqrt{1-\beta}h_{\rm in}^{\rm (\beta)}\label{sup:eqn:aoutb}
\end{equation}

The field then enters the second amplifier, which contributes additional noise to the output field through the noise operator $h_{\rm in}^{\rm (A)}$ \cite{caves1982quantum}:
\begin{align}
a_{\rm out}^{\rm (A)} = &\sqrt{G_{\rm A}}a_{\rm out}^{\rm (\beta)} + \sqrt{G_{\rm A}-1}h_{\rm in}^{\rm (A)}\\
= &\sqrt{G_{\rm A}\beta}\left[\left( \frac{i\kappa_{\rm e}}{\omega-\omega_{\rm r} +i\bar{\kappa} +K(\omega)} -1 \right)a_{\rm in} + \frac{i \sqrt{\kappa_{\rm e}}}{\omega-\omega_{\rm r}+i\bar{\kappa}+K(\omega)} \sum_{j=1}^{N} \frac{\sqrt{\gamma}g_j{d_{\rm in}^{(j)}}^\dagger}{\omega-\omega_j+i\frac{\gamma}{2}} + \frac{i\sqrt{\kappa_{\rm e}\kappa_{\rm i}}}{\omega-\omega_{\rm r}+i\bar{\kappa}+K(\omega)}b_{\rm in}\right]\nonumber\\
&+\sqrt{G_{\rm A}(1-\beta)}h_{\rm in}^{\rm (\beta)} + \sqrt{G_{\rm A}-1}h_{\rm in}^{\rm (A)}
\label{sup:eqn:aouta}
\end{align}

Assuming again that the different fields are uncorrelated, we find the noise at the output of the second amplifier to be:
\begin{equation}
n_{\rm out}^{\rm (A)} = G_{\rm A}\beta\left[R_{\rm in}^+\left(n_{\rm in} + \frac{1}{2}\right) + R_{\rm s}^+\left(n_{\rm s} + \frac{1}{2}\right) +
R_{\rm i}^+\left(n_{\rm i} + \frac{1}{2}\right)\right] + G_{\rm A}(1-\beta)\left(n_{\rm \beta} + \frac{1}{2}\right) + \left(G_{\rm A} - 1\right)\left(n_{\rm A} + \frac{1}{2}\right)\label{sup:eqn:nouta}
\end{equation}
where $R_{\rm in}^+$, $R_{\rm s}^+$ and $R_{\rm i}^+$ are defined in Eqs.~\ref{sup:eqn:rinpm}-\ref{sup:eqn:ripm}, $n_{\rm in}=n_{\rm i}=n_{\rm \beta}=n_0$, and $n_0$ is the thermal population at room temperature ($T_0 = 294$~K), found by setting $T_x=T_{\rm 0}$ in Eq.~\ref{sup:eqn:nxbe}.

With the maser disabled (e.g. by turning the laser off and/or detuning the spin ensemble from the resonator), we have $K(\omega)=0$ and $C(\omega)=0$ and therefore $R_{\rm s}^+ = 0$. It can be shown that in this case $R_{\rm in}^+ + R_{\rm i}^+ = 1$. Eq.~\ref{sup:eqn:nouta} then simplifies to:
\begin{equation}
n_{\rm out,off}^{\rm (A)} = G_{\rm A}\left(n_0 + \frac{1}{2}\right) + \left(G_{\rm A} - 1\right)\left(n_{\rm A} + \frac{1}{2}\right)\label{sup:eqn:noutaoff}
\end{equation}

With the maser amplifier enabled, the output noise is:
\begin{equation}
n_{\rm out,on}^{\rm (A)} = G_{\rm A}R_{\rm in}^+\beta\left[\left(n_0 + \frac{1}{2}\right) + n_{\rm m}\right] + G_{\rm A}(1-\beta)\left(n_0 + \frac{1}{2}\right) + \left(G_{\rm A} - 1\right)\left(n_{\rm A} + \frac{1}{2}\right)\label{sup:eqn:noutaon}
\end{equation}
where we have defined the input-referred noise intrinsic to the maser amplifier $n_{\rm m}$:
\begin{equation}
n_{\rm m} = \frac{R_{\rm s}^+}{R_{\rm in}^+}\left(n_{\rm s} + \frac{1}{2}\right) + \frac{R_{\rm i}^+}{R_{\rm in}^+}\left(n_{\rm 0} + \frac{1}{2}\right)\label{sup:eqn:nm2}
\end{equation}

In our experiments we analyse the noise over a narrow bandwidth at the center of the maser gain profile, such that $R_{\rm in}^+=G_{\rm m}$, where $G_{\rm m}$ is the peak maser gain. We measure the noise power, which is related to the output number of noise photons by $P_{\rm out}^{\rm (A)} = n_{\rm out}^{\rm (A)}\hbar\omega B$, where is $\omega$ the frequency and $B$ is the bandwidth of the measurement. We record the noise with the maser amplifier enabled and with it disabled, then take the ratio of the measured noise powers to calculate the noise gain $G_{\rm n}$:
\begin{equation}
G_{\rm n} = \frac{P_{\rm out,on}^{\rm (A)}}{P_{\rm out,off}^{\rm (A)}}=\frac{n_{\rm out,on}^{\rm (A)}}{n_{\rm out,off}^{\rm (A)}} = \frac{G_{\rm A}G_{\rm m}\beta\left[\left(n_0 + \frac{1}{2}\right) + n_{\rm m}\right] + G_{\rm A}(1-\beta)\left(n_0 + \frac{1}{2}\right) + \left(G_{\rm A} - 1\right)\left(n_{\rm A} + \frac{1}{2}\right)}{G_{\rm A}\left(n_0 + \frac{1}{2}\right) + \left(G_{\rm A} - 1\right)\left(n_{\rm A} + \frac{1}{2}\right)}\label{sup:eqn:gn}
\end{equation}

In our setup $G_{\rm A} \approx 870$ (29.4~dB), we therefore assume that $G_{\rm A}-1 \approx G_{\rm A}$. Rearranging Eq.~\ref{sup:eqn:gn} to solve for the maser noise $n_{\rm m}$, we find:
\begin{equation}
n_{\rm m} \approx \frac{G_{\rm n}\left(n_{\rm 0} + n_{\rm A} + 1\right) - (1-\beta)\left(n_{\rm 0} + \frac{1}{2}\right) - \left(n_{\rm A} + \frac{1}{2}\right)}{G_{\rm m}\beta} - \left(n_{\rm 0} + \frac{1}{2}\right)\label{sup:eqn:nm3}
\end{equation}

We convert the photon numbers to effective temperatures by inverting the Bose-Einstein distribution $T_x = \hbar\omega/[k_{\rm B}\ln{\left(1+1/n_x\right)}]$, with $x=\{\text{m, 0 or A}\}$. Note, for large photon numbers $\ln{\left(1+1/n_x\right)}\approx 1/n_x$ and $T_x \approx \hbar\omega n_x/k_{\rm B}$. In this case we can express Eq.~\ref{sup:eqn:nm3} directly in terms of the noise temperatures:
\begin{equation}
T_{\rm m} \approx \frac{G_{\rm n}\left(T_{\rm 0} + T_{\rm A}\right) - (1-\beta)T_{\rm 0} - T_{\rm A}}{G_{\rm m}\beta} - T_{\rm 0}\label{sup:eqn:tm}
\end{equation}

This approximation is valid in our case, as all temperatures exceed 200~K (i.e. $> 420$ photons at $\omega/2\pi = 9.8$~GHz). We therefore use Eq.~\ref{sup:eqn:tm} to generate the data in Fig.~4a of the main text. Finally, we make a precise measurement of the second-stage amplifier noise temperature $T_{\rm A}$ using a commercial system (Keysight, PNA-X) via the cold-source measurement technique \cite{keysight2024high,pepe2017chip}, as discussed in the Methods section of the main text.

\section{Spin temperature measurements}\label{sup:sec:spin_temp}
The spin polarization induced by optical pumping was determined by performing pulsed ESR experiments at the maser amplification transition. The spins were first polarized in the $m_{\rm s} = 0$ state by a laser pulse of power $P_{\rm L}$ and duration 20~ms. The amount of spin polarization is then estimated by performing a Hahn echo measurement and recording the resulting spin echo signal (see Fig.~4b of the main text). The echo amplitude $A_{\rm e}$ is proportional to the spin polarization $\rho$ \cite{bienfait2016reaching}, which depends on the spin temperature $T_{\rm s}$ as described in Eq.~\ref{sup:eqn:rhotanh}.

\begin{figure}[h]
\includegraphics{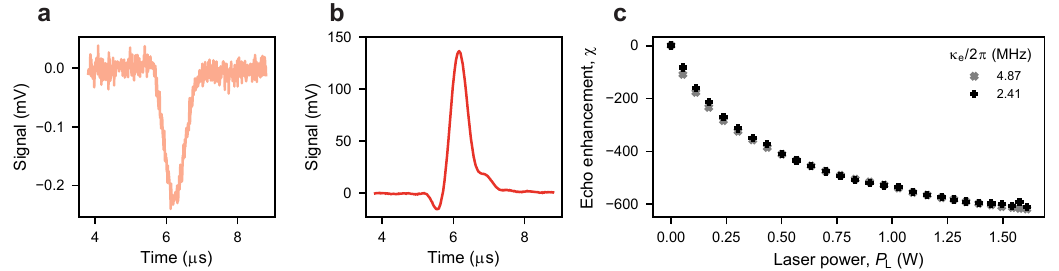}
\caption{\textbf{Hahn echo enhancement with optical pumping.} \textbf{a}, Hahn echo signal without a laser initialization step, where the spin ensemble is in thermal equilibrium at room temperature. \textbf{b}, Hahn echo signal with maximum laser power ($P_{\rm L} = 1.61$~W) applied during the initialization step. \textbf{c}, Enhancement of the spin echoes $\chi$ as a function of laser power $P_{\rm L}$ measured at two different values of the external coupling $\kappa_{\rm e}$.}
\label{sup:fig:Echoes}
\end{figure}

When no optical initialization is performed the spins are in thermal equilibrium at room temperature, i.e. $T_{\rm s} = T_0$. The resulting echo signal is small and plot in Fig.~\ref{sup:fig:Echoes}a. With the laser applied at maximum power, the spin echo grows considerably (Fig.~\ref{sup:fig:Echoes}b). Note, with the laser on the spin population becomes inverted and the echoes appear with the opposite sign compared to the case without optical pumping. We define the echo enhancement $\chi=A_{\rm e}(T_{\rm s})/A_{\rm e}(T_{\rm 0})=\rho(T_{\rm s})/\rho(T_{\rm 0})$, where $T_{\rm s}$ is the effective spin temperature achieved with optical pumping at a laser power of $P_{\rm L}$. We plot $\chi$ as a function of $P_{\rm L}$ in Fig.~\ref{sup:fig:Echoes}c and find a maximum enhancement of $|\chi|\approx 620$.

Using Eq.~\ref{sup:eqn:rhotanh} we find that for a spin transition frequency of $\omega_{\rm s}/2\pi = 9.8$~GHz, the room-temperature spin polarization is $\rho(T_{\rm 0}) = 0.0008$, which allows us to calculate the polarization $\rho(T_{\rm s}) = \mathcal{E}\rho(T_{\rm 0})$ as a function of $P_{\rm L}$ and therefore the spin temperature. We note that due to population inversion, both the polarization and effective spin temperature are negative. In Fig.~4c of the main text, we plot the magnitude of these quantities.

To ensure that the spin echoes observed are not unintentionally amplified by stimulated emission, we perform the measurements in the strongly over-coupled regime with large $\kappa_{\rm e}$. We also take the echo enhancement measurement at two different values of $\kappa_{\rm e}$, with the results compared in Fig.~\ref{sup:fig:Echoes}c. There is no significant quantitative difference between the two measurements, which indicates that the results are not affected by stimulated emission.

\section{Spin-resonator coupling strength} 
In this section we determine the single spin-photon coupling strength ($g_0$) by performing finite element simulations of the mode magnetic field $B_1$ (see Fig.~\ref{sup:fig:DR}a) using CST Studio Suite. Since the NV$^-$ $\left<111\right>$ axis and $B_0$ field are aligned along the x-axis, the relevant component of the field that drives spin rotations is $|B_1^\perp| = \sqrt{|B_1^{\rm y}|^2 + |B_1^{\rm z}|^2}$. 

The single spin coupling strength is given by~\cite{haroche2006exploring, bienfait2016reaching}:
\begin{equation}
g_0 = \gamma_{\rm e}M|\delta B_1^\perp| \label{sup:eqn:g0}
\end{equation} 
where $M=1/\sqrt{2}$ is the matrix element for driving transitions between the $m_{\rm s} = 0$ and $m_{\rm s} = \pm 1$ states, $\gamma_e/2\pi = 28$~GHz/T is the electron spin gyromagnetic ratio and $|\delta B_1^\perp|$ is the perpendicular component of the magnetic field vacuum fluctuations. We therefore need to convert the classical field extracted from our finite element simulation $|B_1^\perp|$ to the vacuum field $|\delta B_1^\perp|$.

A linear oscillating magnetic field is described by the operator $B = \delta B(a+a^\dagger)$. We can relate this to the input field that is used to drive the resonator using input-output theory. Starting with Eq.~\ref{sup:eqn:aw} and ignoring terms that relate to spins (which do not enter the finite-element simulations), we can show:
\begin{equation}
a[\omega] = \frac{2\sqrt{\kappa_{\rm e}}}{2i(\omega_{\rm r} - \omega) +\kappa_{\rm e}+\kappa_{\rm i}}a_{\rm in}[\omega] + \frac{2\sqrt{\kappa_{\rm i}}}{2i(\omega_{\rm r} - \omega) + \kappa_{\rm e}+\kappa_{\rm i}}b_{\rm in}[\omega]
\end{equation}

Taking the solution at resonance $\omega = \omega_{\rm r}$ and making the frequency dependence of the fields implicit, we find:
\begin{equation}
a = \frac{2\sqrt{\kappa_{\rm e}}}{\kappa_{\rm e}+\kappa_{\rm i}}a_{\rm in} + \frac{2\sqrt{\kappa_{\rm i}}}{\kappa_{\rm e}+\kappa_{\rm i}}b_{\rm in}
\end{equation}

The resonator is driven by a coherent microwave tone, putting the system in a coherent state $\left|\alpha\right>$. The classical magnetic field is then described by:
\begin{equation}
\left<B\right> = \delta B\frac{2\sqrt{\kappa_{\rm e}}}{\kappa_{\rm e}+\kappa_{\rm i}}\left(\langle a_{\rm in}\rangle + \langle a_{\rm in}^\dagger\rangle\right) \label{sup:eqn:bexp}
\end{equation}
where we have assumed that the bath is in a thermal state, i.e. $\langle b_{\rm in}\rangle = 0$.

The coherent state $\left|\alpha\right>$ is an eigenstate of $a_{\rm in}$ with complex eigenvalue $\alpha(t) = \alpha e^{-i\omega_{\rm r} t}$, such that $\langle a_{\rm in} \rangle = \alpha(t)$ and $\langle a_{\rm in}^\dagger \rangle = \alpha(t)^*$. Here $|\alpha|^2 = P_{\rm M}/\hbar\omega_{\rm r}$ describes the average number of photons per time incident on the resonator for a microwave drive of power $P_{\rm M}$. From Eq.~\ref{sup:eqn:bexp} and the above definitions, we find:
\begin{equation}
\left<B\right> = 2\delta B\frac{2\sqrt{\kappa_{\rm e}}}{\kappa_{\rm e}+\kappa_{\rm i}}\sqrt{\frac{P_{\rm M}}{\hbar\omega_{\rm r}}}\cos{(\omega_{\rm r} t - \phi)} = 2\delta B\sqrt{\bar{n}}\cos{(\omega_{\rm r} t - \phi)}\label{sup:eqn:bexp2}
\end{equation}
where we define the average number of photons in the resonator $\bar{n}$ as:
\begin{equation}
\bar{n} = \frac{4\kappa_{\rm e}}{\hbar\omega_{\rm r}(\kappa_{\rm e}+\kappa_{\rm i})^2}P_{\rm M} \label{sup:eqn:n_bar}
\end{equation}

In our simulations we extract the field amplitude $|B_1^\perp|$ averaged over the diamond sample. From Eq.~\ref{sup:eqn:bexp2} we can see that this is related to the vacuum fluctuations $|\delta B_1^\perp|$ via $|B_1^\perp| = 2|\delta B_1^\perp|\sqrt{\bar{n}}$. We therefore determine $g_0$ using:
\begin{equation}
g_0 = \frac{\gamma_{\rm e}M|B_1^\perp|}{2\sqrt{\bar{n}}} \label{sup:eqn:g02}
\end{equation} 

The coupling rates $\kappa_{\rm e}$ and $\kappa_{\rm i}$ are found by fitting the simulated reflection response. The power applied in the simulation $P_{\rm M} = 0.5$~W can then be used (Eq.~\ref{sup:eqn:n_bar}) to calculate $\bar{n}$. Putting all parameters into Eq.~\ref{sup:eqn:g02}, we find $g_0/(2\pi) = 0.18$~Hz.

\section{Microwave cooling}\label{sup:sec:cooling}
Here we describe how the microwave cooler output noise temperature, reported in Section~II~D of the main text, is inferred from our data. The microwave cooling measurement setup is identical to the one used in the maser amplifier noise temperature measurements (see Fig.~\ref{sup:fig:NoiseSetup}). We can write down the noise at the output of the second amplifier by replacing the coefficients in Eq.~\ref{sup:eqn:nouta} that are appropriate for amplification ($R_{\rm in}^+$, $R_{\rm s}^+$ and $R_{\rm i}^+$) with those that describe cooling ($R_{\rm in}^-$, $R_{\rm s}^-$ and $R_{\rm i}^-$), as explained in Supplementary Section~\ref{sup:subsec:unified}. The system output noise is therefore given as:
\begin{equation}
n_{\rm out}^{\rm (A)} = G_{\rm A}\beta\left[R_{\rm in}^-\left(n_{\rm in} + \frac{1}{2}\right) + R_{\rm s}^-\left(n_{\rm s} + \frac{1}{2}\right) +
R_{\rm i}^-\left(n_{\rm i} + \frac{1}{2}\right)\right] + G_{\rm A}(1-\beta)\left(n_{\rm \beta} + \frac{1}{2}\right) + \left(G_{\rm A} - 1\right)\left(n_{\rm A} + \frac{1}{2}\right)\label{sup:eqn:noutac}
\end{equation}

We define the microwave cooler output noise as $n_{\rm c} = R_{\rm in}^-\left(n_{\rm in} + 1/2\right) + R_{\rm s}^-\left(n_{\rm s} + 1/2\right) +
R_{\rm i}^-\left(n_{\rm i} + 1/2\right)$ and assume that all baths, apart from the spin bath, are at room temperature, i.e. $n_{\rm in}=n_{\rm i}=n_{\rm \beta}=n_0$. This simplifies the expression for the system output noise to:
\begin{equation}
n_{\rm out,on}^{\rm (A)} = G_{\rm A}\beta n_{\rm c} + G_{\rm A}(1-\beta)\left(n_0 + \frac{1}{2}\right) + \left(G_{\rm A}-1\right)\left(n_{\rm A} + \frac{1}{2}\right)\label{sup:eqn:noutac2}
\end{equation}

With the laser off and spins detuned from the resonator, the cooler output noise becomes $n_{\rm c} = n_0 + 1/2$, and the system output noise $n_{\rm out,off}^{\rm (A)}$ is then given by Eq.~\ref{sup:eqn:noutaoff}. In the experiment we divide the output noise power measured with the spins on resonance and the laser on $P_{\rm out,on}^{\rm (A)}$, with the output power recorded with the spins detuned and the laser off $P_{\rm out,off}^{\rm (A)}$, providing the noise ratio $R_{\rm n}$:
\begin{equation}
R_{\rm n} = \frac{P_{\rm out,on}^{\rm (A)}}{P_{\rm out,off}^{\rm (A)}}=\frac{n_{\rm out,on}^{\rm (A)}}{n_{\rm out,off}^{\rm (A)}} \approx \frac{\beta n_{\rm c} + (1-\beta)\left(n_0 + \frac{1}{2}\right) + \left(n_{\rm A} + \frac{1}{2}\right)}{n_0 + n_{\rm A} + 1}\label{sup:eqn:gnc}
\end{equation}
where we have used the approximation $G_{\rm A} - 1 \approx G_A$. Solving for $n_{\rm c}$:
\begin{equation}
n_{\rm c} = \frac{R_{\rm n}\left(n_0 + n_{\rm A} + 1\right) - (1-\beta)\left(n_0 + \frac{1}{2}\right) - \left(n_{\rm A} + \frac{1}{2}\right)}{\beta}
\end{equation}
If we assume all photon numbers are large ($\gg 1$), then we can write this directly in terms of temperatures:
\begin{equation}
T_{\rm c} \approx \frac{R_{\rm n}\left(T_0 + T_{\rm A}\right) - (1-\beta)T_0 - T_{\rm A}}{\beta}
\end{equation}

\section{ESR spectroscopy experiments}\label{sup:sec:ESR}
This section details measurements of the spin resonance spectra as well as spin relaxation ($T_1$) and coherence ($T_2$) times. All experiments were performed using the pulsed ESR setup detailed in Fig.~\ref{sup:fig:ESRSetup}.

\begin{figure}[h]
\includegraphics{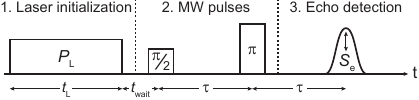}
\caption{\textbf{Hahn echo pulse sequence.} The sequence begins by initializing the spin ensemble with a 20~ms long laser pulse at an optical power $P_{\rm L}$. After a time $t_{\rm wait}$ we perform a two-pulse Hahn echo experiment and record the area of the resulting spin echo $S_{\rm e}$.}
\label{sup:fig:T1_T2_pulse_sequence}
\end{figure}

\subsection{ESR spectra}\label{sup:subsec:spectra}
\begin{figure}[h]
\includegraphics{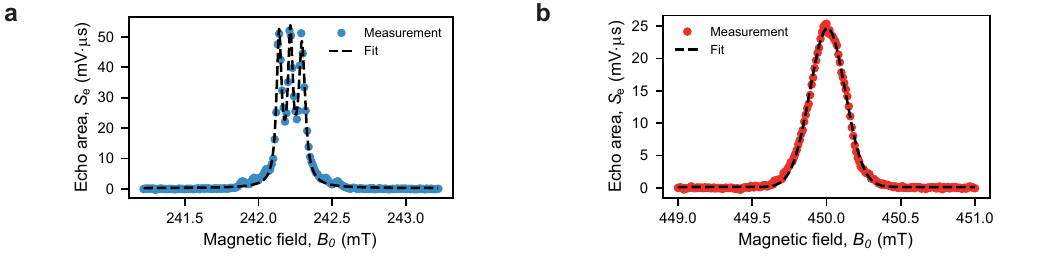}
\caption{\textbf{ESR spectra.} \textbf{a}, Spectra of the cooling transition, fitted with a triple Lorentzian function. The full width at half maximum of the $m_I=-1$, $m_I=0$ and $m_I=+1$ hyperfine peaks are found to be $41(2)$~$\mu$T, $39(1)$~$\mu$T and $52(2)$~$\mu$T, respectively. \textbf{b}, Spectra of the amplification transition, fitted to a single Gaussian peak. A standard deviation of 119(1)~$\mu$T is extracted from the fit.}
\label{sup:fig:Linewidths}
\end{figure}

We measured the ESR spectra of the NV$^-$ spins at both the cooling and amplification transitions using an echo-detected field sweep. We apply the Hahn echo sequence shown in Fig.~\ref{sup:fig:T1_T2_pulse_sequence} while stepping the magnetic field and recording the detected spin echo. The signal is integrated to determine the echo area $S_{\rm e}$ and then plot against $B_0$. At $B_0 \approx 242$~mT we observe the cooling spin transition (Fig~\ref{sup:fig:Linewidths}a), where we can partially resolve the hyperfine splitting caused the NV$^-$ electronic spin interacting with the $^{14}$N nuclear spins ($I = 1$). Each of these hyperfine peaks has a full width at half maximum (FWHM) of approximately 45~$\mu$T, indicating the high quality of the CVD diamond sample. At $B_0 \approx 520$~mT we find the amplification transition (Fig.~\ref{sup:fig:Linewidths}b), where we observe that the hyperfine peaks are inhomogenously broadened, likely by the $B_0$ magnetic field (see main text, Section~IV~B). The spectrum is fit well by a Gaussian function, with an extracted standard deviation of $119(1)~\mu$T.

\subsection{Spin energy relaxation ($T_1$) and coherence ($T_2$) times}\label{sup:subsec:T1T2}
\begin{figure}[h]
\includegraphics{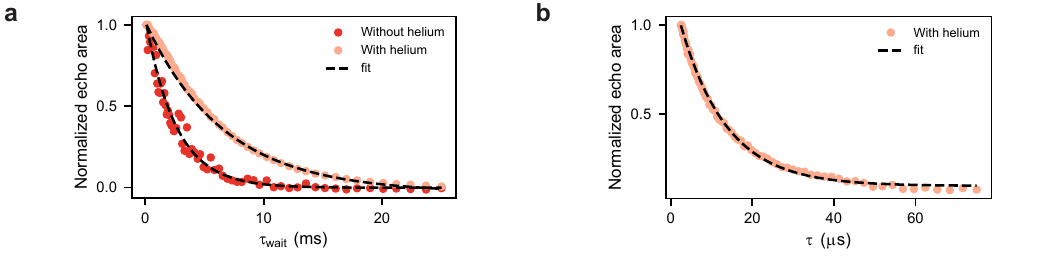}
\caption{\textbf{Spin energy relaxation ($T_1$) and coherence ($T_2$) time measurements.} \textbf{a}, $T_1$ measurements with and without helium gas flowing over the diamond and resonator. \textbf{b}, $T_2$ measurement with helium gas applied. All fits in panels a and b are simple exponential decays.}
\label{sup:fig:T1_T2}
\end{figure}

We measure the $T_1$ and $T_2$ times at the maser amplification transition using the sequence in Fig.~\ref{sup:fig:T1_T2_pulse_sequence}. To measure $T_1$ we keep the delay in the Hahn echo sequence fixed at $\tau = 5~\mu$s and vary the time $t_{\rm wait}$ between the laser initialization pulse and the start of the Hahn echo. In Fig.~\ref{sup:fig:T1_T2}a we plot the normalized echo signal versus $t_{\rm wait}$ for two measurements, one with helium gas flowing over the resonator/diamond assembly and one without helium flowing. With helium we find a $T_1 = 6.65(10)$~ms, which drops to $T_1 = 2.57(12)$~ms without helium. This emphasises the importance of ensuring that the components are properly thermalized during operation.

To measure $T_2$, we fix the wait time to $t_{\rm wait} = 100$~$\mu$s and vary the delay between the pulses $\tau$ in the Hahn echo sequence. From the decay of the normalized echo signal shown in Fig.~\ref{sup:fig:T1_T2}b, we extract a $T_2 = 11.32(17)~\mu$s. This value is maintained even without a flow of helium gas, indicating that the current amount of sample heating has little impact on spin coherence.

\bibliographystyle{unsrt}
\bibliography{bib}